\documentclass[review,3p,times]{elsarticle}
\usepackage{epstopdf}
\usepackage{amssymb}
\usepackage{hyperref}
\usepackage{amsmath}

\newcommand{\psibar}{\bar \psi}
\DeclareMathOperator* {\GoesToWhen}{\longrightarrow}
\newcommand{\rot}{\mathop{\rm rot}\nolimits}
\newcommand{\Det}{\mathop{\rm Det}\nolimits}
\newcommand {\Veff} {\ensuremath{V_{\rm eff}}}
\newcommand{\e}{\mathop{\rm e}\nolimits}
\newcommand {\Vo}{\ensuremath{V_{(0)}}}
\newcommand {\VL}{\ensuremath{V_{(L)}}}
\newcommand {\VmT}{\ensuremath{V_{(\mu T)}}}
\DeclareMathOperator {\Li}{Li}
\newcommand{\ua}{\uparrow}
\newcommand{\da}{\downarrow}

\journal{Annals of Physics}
\begin{document}

\begin{frontmatter}

\title{Phase transitions in hexagonal,
  graphene-like lattice sheets and nanotubes
under the influence of external conditions}

\author[AffilationEbert]{D. Ebert}

\author[AffilationKlimenko]{K.G. Klimenko}

\author[AffilationMSU]{\corref{mycorrespondingauthor}P.B. Kolmakov}
\cortext[mycorrespondingauthor]{Corresponding author}
\ead{pavel.b.kolmakov@yandex.ru}

\author[AffilationMSU]{V.Ch. Zhukovsky}
\address[AffilationEbert]{Institute of Physics,
Humboldt-University Berlin, 12489 Berlin, Germany}
\address[AffilationKlimenko]{State Research Center of Russian Federation --
Institute for High Energy Physics, NRC "Kurchatov Institute",
142281, Protvino, Moscow Region, Russia}
\address[AffilationMSU]{Faculty
of Physics, Moscow State University, 119991, Moscow, Russia}

\begin{abstract}
In this paper we consider a class of (2+1)D schematic models with
four-fermion interactions that are effectively used in studying
condensed-matter systems with planar crystal structure,
and especially graphene.
Symmetry breaking in these models occurs due to a possible
appearance of condensates. Special attention is paid to the symmetry
properties  of the appearing condensates in the framework of
discrete chiral and
$\mathcal C$, $\mathcal P$ and $\mathcal T$ transformations.
Moreover, boundary conditions corresponding to carbon nanotubes are
considered and  their relations with the effect of an applied
external magnetic field are studied. To this end we calculated the
effective potential for the nanotube model including effects of
finite temperature, density and an external magnetic field. As an
illustration we made numerical calculations of the chiral symmetry
properties in a simpler Gross--Neveu model with only one condensate
taken into account. We also investigated the phase structure of the
nanotube model under the influence of the Aharonov--Bohm effect and
demonstrated that there is a nontrivial relation between the
magnitude of the  Aharonov--Bohm phase, compactification of the
spatial dimension and thermal restoration of the originally broken
chiral symmetry.
\end{abstract}

\begin{keyword}
effective Lagrangian, phase transitions, carbon nanotubes, Zeeman effect, Aharonov--Bohm effect
\PACS 73.22.Gk \sep 11.10.Wx \sep 71.70.Ej

\end{keyword}

\end{frontmatter}

\section{Introduction}

It is a well-known fact that relativistic quantum field theory
provides a powerful tool for the description of low-energy
excitations in condensed-matter physics \cite{n1}. Examples are the
field theoretic description of low-energy electron states in
polymers \cite{n2a,n2b,n2c} or the recent quasirelativistic
treatment of electrons in planar systems like graphene, a single
layer of graphite \cite{n3}. Recall that in the case of graphene,
the original nonrelativistic tight-binding model for electrons on a
hexagonal ``honeycomb'' lattice admits a low-momentum expansion
around the two inequivalent ``Dirac points'', the corners (valleys)
of the first Brillouin zone, which leads to a linear dispersion law
for low-energy fermion excitations, closely resembling that of
massless relativistic Dirac fermions \cite{n3,n4}. Combining the two
valley degrees of freedom with the two sublattice (pseudospin)
degrees of freedom of electrons of carbon atoms, leads in a natural
way to a reducible four-component Dirac spinor description in
D=(2+1) dimensions. It is just this property which allows for the
introduction of a chiral $\gamma_5$-matrix and the use of a chiral
(Weyl) representation of Dirac matrices \cite{n5}. In the continuum limit, the free Dirac Lagrangian of graphene develops an emergent
chiral "valley-sublattice" $U(2)_{vs}$ symmetry, which, when
considering ``multilayer'' graphene with $N_{\rm f}$ flavors, is further
enlarged to a chiral $U(2N_{\rm f})$ symmetry. There arises then the
important question, whether the inclusion of fermion interactions
can lead to a dynamical breakdown of chiral symmetry with an
associated dynamical fermion mass generation and a
``semimetal-insulator'' phase transition.

The phenomenon of a dynamical generation of a fermion mass on the
basis of a generic four-fermion interaction is well-known for strong
interactions since the time, when Nambu and Jona-Lasinio (NJL)
\cite{n6} generalized the BCS-Bogoliubov theory \cite{n7,n8} of
superconductivity to a relativistic model with dynamical breaking of
a continuous $\gamma_5$-symmetry. Later on, QCD-motivated NJL-type
of models were shown to successfully describe the low-energy meson
spectrum of quantum chromodynamics (QCD) \cite{n9}. Similar types of
four-fermion models with a discrete $\gamma_5$-symmetry have also
been considered in lower dimensions D=(1+1) by Gross and Neveu (GN)
\cite{n10}, where the four-fermion theory is renormalizable and
asymptotic free, or for D=(2+1) in refs. \cite{n11,24}. In the latter
case, the model is perturbatively nonrenormalizable but becomes
renormalizable in the $1/N_{\rm f}$ expansion \cite{28}.

Generally, four-fermion models provide a useful effective low-energy
description of an underlying relativistic fundamental theory. This
fact makes it further interesting to investigate their modifications
under the influence of external conditions, such as temperature,
chemical potential, external magnetic fields etc.
\cite{n12,n13,n14,n15}. Note, on the other hand, that in
condensed-matter physics such models are meant to be effective from
the very beginning. Non-renormalizability makes here no additional
problem due to the natural cutoff in the ultraviolet momentum region
provided by the finite spacing between the elements of the polymer
lattice. Obviously, one may expect that local four-fermion
interactions play also an important role for the generation of a
dynamical mass gap and quantum phase transitions in graphene
\cite{n16,n17,n18}. Let us refer in this context also to the interesting
investigations based on Schwinger-Dyson equations \cite{n19,n20},
renormalization group flow equations \cite{n16} and functional
renormalization group methods \cite{n21,n22}.

    The main aim of this paper is to continue investigations
based on approximative local four-fermion interactions of fermions
in a graphene-like hexagonal lattice. In particular, we shall apply
the method of the effective potential and the mean-field approach to
describe fermionic quasiparticles and excitonic bound states for
graphene-like lattice sheets and nanotubes. Moreover, we shall
investigate phase transitions under external conditions like
temperature, chemical potential and Aharonov-Bohm (AB) magnetic
fields \cite{32}.

    The paper is organized as follows. In Sect. 2 we first review the
effective low-energy model of $N_{\rm f}$ non-interacting fermion species
(flavors) living on a planar honeycomb lattice. Particular attention
is paid to the emergent chiral symmetry $U(2N_{\rm f})$. Next, we will
consider possible $U(2N_{\rm f})$-invariant effective four-fermion
interactions obtained by a contact approximation of the
instantaneous Coulomb potential using the ``braneworld'' or
``reduced QED scenario'' proposed in ref. \cite{n23}. In this
approach fermions are localized on a 2D brane and move with Fermi
velocity $v_{\rm F}$, whereas the electromagnetic gauge field propagates
in the 3D bulk with speed of light $c$. In graphene one has $v_{\rm F}/c
\approx 1/300$, so that the effective fine-structure constant is
$\alpha_{\rm eff}=\alpha c/v_{\rm F} \approx 2$ ($\alpha=1/137$), providing us
with an interesting strong coupling theory. Next, the obtained
four-fermion Coulomb-based interaction is projected by a
Fierz-transformation into the fermion-hole channels, where exciton
bound states can occur. This results in an effective NJL-type of
interactions which keeps the global $U(2N_{\rm f})$ symmetry intact.
However, one cannot postulate exact chiral $U(2N_{\rm f})$ symmetry from
the very beginning. The reason is that chiral symmetry arises only
in the continuum limit and is not exact in the tight-binding lattice
Hamiltonian \cite{n24,n25}. This requires to admit additional small
on-side repulsive interaction terms which break the symmetry
explicitly \cite{n19,n26}. By this reason, we finally omit all
possible symmetry constraints between four-fermion couplings and
start with a general schematic GN-type of model, considered in
earlier papers \cite{n13,n27} for (2+1)D QED and QCD. Although not
directly related to graphene, their methods turn out to be useful
for our mathematical investigations. Finally, we also quote the
symmetry breaking properties of fermion condensates concerning
discrete $\mathcal P$, $\mathcal C$, $\mathcal T$, $\gamma_5$ and
$\gamma_3$ transformations.

     In Sect. 3 we perform the path-integral derivation of the
effective potential in the large $N_{\rm f}$ (mean field) approximation.
The global minimum point of the effective potential then determines
the fermion mass gap. In addition, we determine the exciton spectrum
from the two-point 1PI Green functions (inverse propagators) of
fluctuating exciton fields. Finally, for possible applications to
nanotubes, we compactify one spatial direction rolling up the
honeycomb lattice to a cylinder. In the resulting nanotubes we shall
take into account the effects of finite temperature, chemical
potential and of magnetic AB fields. The influence of the magnetic
AB-effect on similar compactified fermion systems was recently
studied in refs. \cite{n28,n29,n30,Ebert} (see also the 5D model
\cite{n31}), and earlier in ref. \cite{n32} and in physics of carbon
nanotubes in refs. \cite{n33,n34,n35}.

     In Sect. 4 we numerically investigate chiral phase transitions in
nanotubes in the presence of finite temperature, chemical potential
and the magnetic AB-phase $\phi$. In particular, we present phase
portraits in the $(T,\mu)$, $(\beta,L)$ and $(\phi,\beta)$
planes, where $\beta=1/T$ and $L=2\pi R$ (R is the cylinder radius).

     Sect. 5 contains our summary and conclusions. Technical details of
the Fierz transformation, of the investigation of the phase structure of the considered general schematic GN-type of model and of fermion loop calculations for exciton propagators are relegated to three appendices.

\section{Effective low-energy model \label{sec2}}

\subsection{Non-interacting fermions on a planar honeycomb lattice}

\begin{figure}
\begin{center}
\includegraphics[width=400pt]{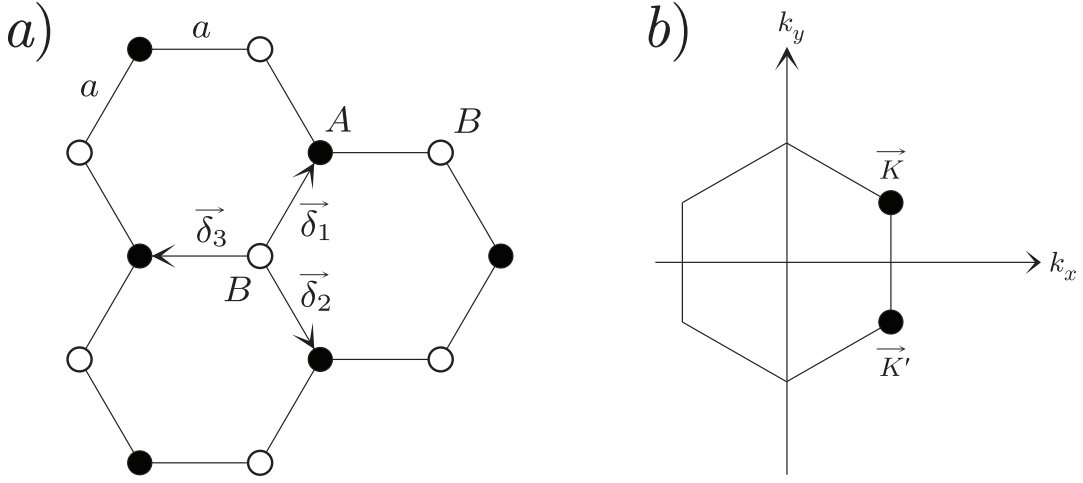}
\caption{\it{a.) Hexagonal honeycomb lattice with two
interpenetrating triangular lattices of A and B sites. $\vec \delta_i,\,i=1,2,3$ are the
nearest neighbor vectors. b.) Corresponding Brillouin zone: the Dirac cones of the fermion
spectrum are located at the $K$ and $K'$ points.}}\label{Fig.1}
\end{center}
\end{figure}
It is well known that the hopping of fermions living on a
graphene-like hexagonal "honeycomb" lattice can be described by the
following tight binding Hamiltonian \cite{1} (for reviews see refs.
\cite{n3,n5})
\begin{equation}
H_0=-t\sum\limits_{\vec r\in B}\sum\limits_{i=1,2,3}
\left[\psi^{+Aa}(\vec{r}+\vec \delta_i)\psi^{Ba}(\vec r)+h.c.\right]
.\label{2.1}\end{equation} Here $t$ is the nearest neighbor hopping
parameter, $\psi^{+Aa}$ and $\psi^{Ba}$ are Fermi field operators
belonging to triangular sublattices with A and B sites, and $\vec
\delta_i,\,i=1,2,3$ are three vectors directed from a B site to
three nearest neighbor A sites. They are given by (Fig.
\ref{Fig.1}{\it a}) $\vec\delta_1=\frac{a}{2}(1,\sqrt{3})$,
$\vec\delta_2=\frac{a}{2}(1,-\sqrt{3})$, $\vec\delta_3=-a(1,0)$ with
$a$ being the distance between lattice sites.

For later use of a $1/N_{\rm f}$ expansion, we consider here the
"multilayer" case of $N_{\rm f}=2N$ degenerate fermion species (flavors)
of real spin $\uparrow$ and $\downarrow$, living on $N$ hexagonal
monolayers which are described by fields with a flavor index
$a=(1,...,N_{\rm f}=2N)$.
The monolayer case $N_{\rm f}=2$ corresponds to a
fermion with two spin projections. Note that repeated indices $a$ in
eq. (\ref{2.1}) have to be summed over and will be generally
omitted.

In momentum representation, the Hamiltonian becomes diagonal,
\begin{equation}
H_0=\sum\limits_{\vec k} \left[ \Phi(\vec k)\psi^{+Aa}(\vec
k)\psi^{Ba}(\vec k)+h.c. \right],~~~~\Phi(\vec
k)=-t\sum\limits_{\vec\delta_i}e^{-i\vec k \vec
\delta_i}.\label{2.2}
\end{equation}
The energy bands, derived from this Hamiltonian are \cite{n4}
\begin{equation}
\mathcal{E}_{\pm}(\vec k)=\pm\left|\Phi(\vec k)\right|
,\label{2.3}\end{equation} where the $+/-$ signs refer to the
upper/lower bands. It is important that there exist two inequivalent
points $K$ and $K'$ ("Dirac points") at the corners of the first
Brillouin zone, where $\mathcal{E}_{\pm}(\overrightarrow
K;\overrightarrow{K}')=0$. Their positions in the momentum space is
given by (Fig. \ref{Fig.1}{\it b})
\begin{equation}
\overrightarrow
K=\left(\frac{2\pi}{3a},\frac{2\pi}{3\sqrt{3}a}\right),\;
\overrightarrow
K'=\left(\frac{2\pi}{3a},-\frac{2\pi}{3\sqrt{3}a}\right).
\label{2.4}\end{equation}
Performing a low-momentum expansion of the
energy spectrum (\ref{2.3}) around the Dirac points, $\vec
k=\overrightarrow K(\overrightarrow K')+\vec p$, with $\left|\vec
p\right|a<1$, one obtains the famous linear dispersion law $\mathcal
E_\pm=\pm v_{\rm F}\left|\vec
  p\right|$
for massless quasiparticles on the honeycomb lattice \footnote{ In
this paper we use natural units $\hbar=c=k_{\rm B}=1$, where $k_{\rm
B}$ is the Boltzmann constant. Obviously, the Fermi velocity $v_{\rm
F}$ is then also dimensionless. }. Here $v_{\rm F}=\frac{3}{2}ta$ is
the Fermi velocity with $a$ being the lattice spacing. Transforming
the low-momentum expansion of the Hamiltonian (\ref{2.2}) back to
configuration space gives in the continuum limit the Dirac-like free
Hamiltonian (see e.g. refs. \cite{n3,5})
\begin{equation}
H_0=-\sum\limits_{\eta=\pm 1}\sum\limits_{a=1}^{N_{\rm f}}\int d^2 x
\psi_\eta^{+a}(\vec r) \left[v_{\rm F}\tau^1i\partial_x+\eta v_{\rm
F}\tau^2 i\partial_y\right] \psi_\eta^a(\vec r),\label{2.5}
\end{equation}
where $\tau^i$ are $2\times2$ Pauli matrices. The Fermi operators
$\psi_\eta^a(\vec r)$ ($\vec r=(x,y)$) are two-spinors
\begin{equation}
\psi_\eta^a(\vec r)=
\begin{pmatrix}
\psi_\eta^{Aa}\\
\psi_\eta^{Ba}
\end{pmatrix}
\label{2.6}\end{equation} with indices A, B denoting sublattice
("pseudospin") degrees of freedom, and the subscript $\eta=\pm1$
("valley index") stands for the two Dirac points $K$, $K'$
corresponding to valleys of the energy spectrum at the corners of
the first Brillouin zone.

In what follows we shall be interested in the spontaneous breakdown
of chiral symmetry which requires the existence of a chiral
$\gamma^5$-matrix. Such a matrix can only be  obtained by using a
reducible $4\times4$ representation of Dirac matrices.

Following Gusynin at al. \cite{n5}, it is convenient to use the
reducible chiral (Weyl) representation
\begin{equation}
\gamma^0=
\begin{pmatrix}
0 & {\rm I}_2\\
{\rm I}_2 & 0 \\
\end{pmatrix},\;
\gamma^1=
\begin{pmatrix}
0 & -\tau^1\\
\tau^1 & 0 \\
\end{pmatrix},\;
\gamma^2=
\begin{pmatrix}
0 & -\tau^2\\
\tau^2 & 0 \\
\end{pmatrix}
\label{2.7}\end{equation}
with ${\rm I}_2$ being the $2\times2$ unit
matrix. There exist two more $4\times4$ matrices which anticommute
with all $\gamma^\mu$, $\mu=0,1,2$ and with each other
\begin{equation}
\gamma^3=
\begin{pmatrix}
0 & -\tau^3\\
\tau^3 & 0 \\
\end{pmatrix},\;
\gamma^5=
\begin{pmatrix}
{\rm I}_2 & 0\\
0 & -{\rm I}_2
\end{pmatrix}
,\label{2.8a}\end{equation} as well as their combination
\begin{equation}
\gamma^{35}=\frac{1}{2}\left[\gamma^3,\gamma^5\right]=
\begin{pmatrix}
0 & \tau^3\\
\tau^3 & 0
\end{pmatrix},
\label{2.8b}\end{equation}
which commutes with all $\gamma^\mu$, but
anticommutes with $\gamma^3$ and $\gamma^5$. Note that
($\mu=0,1,2,3$)
\begin{equation}
\left\{\gamma^\mu,\gamma^\nu\right\}=2g^{\mu\nu}{\rm I}_4,\;
g^{\mu\nu}={\rm diag}(1,-1,-1,-1), \label{2.9}\end{equation} where
${\rm I}_4$ is the $4\times4$ unit matrix. Let us now replace the
operator-valued fields by 4-spinor Grassmann fields
\begin{equation}
\psi^t=\left(\psi_{K}^{Aa},\psi_{K}^{Ba},-i\psi_{K'}^{Ba},i\psi_{K'}^{Aa}\right)
,\label{2.10}\end{equation} where $t$ stands for the transposition
operation.

By using the notations $\psi_\eta^{A,B}\equiv\psi_{K,K'}^{A,B}$ for
$\eta=\pm1$, one can reexpress the Hamiltonian (\ref{2.5}) in a
convenient 4-spinor notation. With $\psibar^a=\psi^{+a}\gamma^0$,
one arrives at the effective free low-energy Lagrangian
\begin{equation}
L_0=\psibar\left[i\gamma^0\partial_0+ iv_{\rm F}\gamma^1\partial_x+
iv_{\rm F}\gamma^2\partial_y \right]\psi= \psibar
i\gamma^\mu\tilde\partial_\mu\psi, \label{2.11}\end{equation} where
$\tilde\partial_\mu=(\partial_0,v_{\rm F}\vec\nabla)$,
using
notations $x^0=t$, $\partial_0=\partial_t$, and implicit summation
over the flavor index $a$ is understood.

It is illuminative to introduce also chiral projection operators\\
$\mathcal P_\pm=\frac{1}{2}(1\pm\gamma^5)$ and "right" and "left"
spinors $\psi_\pm=\mathcal P_\pm \psi$, i.e.
\begin{equation}
\psi_+=
\begin{pmatrix}
\psi_K^{Aa}\\
\psi_K^{Ba}\\
0\\
0
\end{pmatrix},\;\;
\psi_-=
\begin{pmatrix}
0\\
0\\
-i\psi_{K'}^{Ba}\\
i\psi_{K'}^{Aa}
\end{pmatrix},
\label{2.12}\end{equation} so that
\begin{equation}
\gamma^5\psi_\pm=\pm\psi_\pm. \label{2.13}\end{equation} In the
chiral representation of Dirac matrices, the fermion excitations at
the two distant Dirac points $K$, $K'$ ($\eta=\pm1$), corresponding
to spinors $\psi_\pm$, thus turn out to be states of definite
chirality eigenvalues $\pm1$. Obviously, the latter coincide with
the values of the valley index $\eta=\pm1$ \footnote{ Note that the
chirality eigenvalues coincide with the eigenvalues of the helicity
operator $\Lambda_p=\frac{\vec p\cdot\vec \Sigma}{|\vec p|}$,
where $\vec \Sigma=\text{diag}(\vec\tau,\vec\tau)$ is the
pseudospin.}.

\subsection{Symmetry properties}

It is straightforward to see that the matrices $\gamma^3$,
$\gamma^5$ and $\gamma^{35}$ together with the $4\times4$ unit
matrix ${\rm I}_4$ are generators of an emergent global continuous
"valley-sublattice" symmetry $U(2)_{vs}=U(1)_{vs}\times SU(2)_{vs}$
\cite{n5}. Indeed, it is easy to see that the three generators
\begin{equation}
t^1=\frac{1}{2}i\gamma^3,\;t^2=\frac{1}{2}\gamma^5,\;t^3=\frac{1}{2}\gamma^{35}
\label{2.14}\end{equation} commute with the free Lagrangian
(\ref{2.11}) and satisfy the $SU(2)$ algebra
\begin{equation}
\left[t^i,t^j\right]=i\varepsilon_{ijk}t^k
,\label{2.15}\end{equation} where $\varepsilon_{ijk}$ is the
Levi-Civita symbol. Moreover, let us introduce
the $U(1)$-generator $t^0=\frac{1}{2}{\rm
I}_4$. Then we have the normalization ${\rm tr}\, t^i
t^j=\delta^{ij}$, where $i,j=0,..,3$. Clearly, one may also consider
special continuous $U(1)_{t^k}$ transformations ($k=0,..,3$),
related to the generators $t^k$ in eq. (\ref{2.14}) and to the
generator $t^0$,
\begin{equation}
U(1)_{t^k}:\;\psi\rightarrow e^{i\alpha_k
t^k}\psi,\;\bar{\psi}\rightarrow\bar{\psi}e^{-is_k\alpha_k t^k}
,\label{2.16} \end{equation} where $s_k=-1$ for $k=1,2$ and $s_k=1$
for $k=0,3$. In addition to the above $U(2)_{vs}$ valley-sublattice
symmetry, the Lagrangian (\ref{2.11}) exhibits invariance under the
global group $U(N_{\rm f})$ of flavor symmetry. In fact, it is invariant
under the larger group $U(2N_{\rm f})$, spanned by the generators given by
the direct products
\begin{equation}
t^i\otimes\frac{\lambda^\alpha}{2}\otimes\frac{\sigma^m}{2}~~~~~~~
i=(0,..,3),\, \alpha=(0,..,N^2-1),\, m=(0,..,3),\,\,\,N_{\rm f}=2N,
\label{2.17}\end{equation} with
$\lambda^\alpha\;(\alpha=0,..,N^2-1)$ being generalized Gell-Mann
matrices of $U(N)$ with ${\rm
tr}\,\lambda^\alpha\lambda^\beta=2\delta^{\alpha\beta}$,
$\lambda^0=\sqrt{\frac{2}{N}}{\rm I}_N$,  and $\sigma^m$ are the
Pauli spin matrices ($\sigma^0={\rm I}_2$) of the spin rotation
group $U(2)_s$.

For later use, let us also quote the transformation laws of
4-spinors under the discrete symmetries: inversion of $x$-coordinate
$\mathcal P$, charge conjugation $\mathcal C$, and time reversal
$\mathcal T$,
\begin{equation} \begin{split}
\psi(x^0,x,y)&\stackrel{\mathcal
  P}{\longrightarrow}i\gamma^1\gamma^5\psi(x^0,-x,y),\\
\psi(x^0,\vec r)&\stackrel{\mathcal
  C}{\longrightarrow}\gamma^1\psibar^{t}(x^0,\vec
r),\label{2.18}\\
\psi(x^0,\vec r)&\stackrel{\mathcal
  T}{\longrightarrow}i\sigma^2\gamma^1\gamma^5
\psi(-x^0,\vec r)
\end{split}
\end{equation}
where $\sigma^2$ acts on the spin indices of the spinor.

\subsection{Effective four-fermion interactions}
\subsubsection{Contact approximation of electromagnetic interactions}

As usual, electromagnetic interactions between quasiparticles are
introduced into the free Lagrangian $L_0$ of eq. (\ref{2.11}) by
covariant derivatives
$\widetilde{\partial}_\mu\rightarrow\widetilde{D}_\mu=(\partial_0-ieA_0,v_{\rm
  F}(\vec \nabla+ie\vec A))$. Let us consider here the "braneworld" or
"reduced" QED-scenario proposed in ref. \cite{n23} and start with
the Dirac-Maxwell action \footnote{ The electromagnetic vector
potential $\vec A$ can be introduced in the lattice Hamiltonian
(\ref{2.1}) with the Peierls' substitution, i.e. by introducing the
phase factor $\exp(-ie {\vec\delta}_i \cdot\vec A)$ into the hopping
term \cite{n5}. In a similar way, the scalar potential can be
obtained from changing the next-to-nearest hoppings \cite{n3,7,5}.}
\begin{equation}
S=\int{d^3x\psibar i \gamma^\mu\widetilde{D}_\mu\psi
-\frac{
\varepsilon_0
}{4}\sum \limits_{\mu,\nu=(0,\dots,3)} \int d^4x
F_{\mu\nu}F^{\mu\nu}} .\label{2.19}\end{equation}
Here the fermionic
quasiparticles run in the (2+1)-dimensional space-time
$x^{(3)}=(x^0,x^1,x^2)$ with Fermi velocity $v_{\rm F}$, while the
$U(1)$ gauge field $A^\mu$ propagates in (3+1)-dimensional bulk
space-time $x^{(4)}=(x^0,x^1,x^2,x^3)$ with the speed of light
$c(=1)$. Notice that the gauge field appearing in the covariant derivative 
$\widetilde{\partial}_\mu\rightarrow\widetilde{D}_\mu$ is taken on the
plane $x^3=0$.

The gauge coupling constant $e$
($-e<0$)
is the electric charge for the vacuum-suspended honeycomb lattice,
and $\varepsilon_0$ is the dielectric constant of the vacuum.
If the
layer is placed on a substrate, the interaction strength is screened
by the factor $2/(1+\varepsilon)$ with $\varepsilon$ being the
dielectric constant of the substrate \cite{8}. Let us slightly
rewrite the action (\ref{2.19}) in a form making the coupling of the
gauge field to fermion charge and current densities $\rho$, $\vec j$
explicit,
\begin{equation}
S=-\frac{
\varepsilon_0
}{4}\sum\limits_{\mu,\nu=(0,..,3)}\int{ d^4x
F_{\mu\nu}F^{\mu\nu}}+ \int{d^3x L_0}+ \int{d^3x\left[A_0\rho-\vec
A\cdot\vec j\right]} ,\label{2.19'}\end{equation}
with $L_0$ given
in eq. (\ref{2.11}), and
\begin{equation}
\rho=e\psibar \gamma^0\psi,~~j^1=e v_{\rm F} \psibar \gamma^1
\psi, ~~j^2=e v_{\rm F} \psibar \gamma^2 \psi, ~~j^3=0.
\label{2.20}\end{equation}
As has been demonstrated in  ref. \cite{n23}, the action $S$
can be further rewritten by introducing a new (2+1)-dimensional
``brane'' gauge field $A_{\mu}(x^0,x^1,x^2)$, ~~$(\mu=0,1,2)$.
The resulting effective action then takes the form
\begin{equation}
S=\int{d^3x\left[
-\sum\limits_{\mu,\nu=(0,..,2)}
\frac{\varepsilon_0}{2}F_{\mu\nu}
\frac{1}{\sqrt{-\partial^2}}
F^{\mu\nu}+L_0+A_0\rho-\vec
A\cdot\vec j+{\rm gauge\, terms} \right]} \label{2.21}\end{equation}
with a nonlocal kinetic term of the gauge field.
Finally, let us consider the partition function for the action
(\ref{2.21}),
\begin{equation}
Z=\int{D\psi D\psibar D_\mu[A_\mu]\exp[iS]}
,\label{2.22}\end{equation} where the path-integral measure
$D_\mu[A_\mu]$ includes the gauge-fixing terms. For the following
discussion it turns out convenient to reintroduce, for a moment,
again the speed of light $c$. By integrating out the gauge field and
neglecting relativistic corrections of order $(v_{\rm F}/c)^2$ (for
graphene we have
 $v_{\rm F}/c\sim
1/300$), arising from currents $\vec j$, one obtains the following
expression for the action containing Coulomb interactions of
fermions on the lattice plane
\begin{equation}
S=S_0-\frac{v_{\rm F}}{2c}\int{d^{(3)}x'}\int{d^{(3)}x}
\left[\psibar(x^0,\vec r)\gamma^0\psi(x^0,\vec r) \right]
U_{0}^{C}(x^0-x'^0,\left|\vec r-\vec r\,'\right|)
\left[\psibar(x'^0,\vec r\,')\gamma^0\psi(x'^0,\vec r\,') \right]
.\label{2.23}\end{equation} Here $U_{0}^{C}$ is the bare
instantaneous Coulomb potential which takes the form
\begin{equation}
U_{0}^{C}(x^0,|\vec r|)= \frac{e^2\delta(x^0)}{2\varepsilon_0
v_{\rm F}}\int{\frac{d^2k}{(2\pi)^2}} \exp(i\vec k \vec r)\frac{1}{|\vec k|}={\alpha}\left(\frac{c}{v_{\rm
F}}\right)\frac{\delta(x^0)}{|\vec r|} ,\label{2.24}\end{equation}
where $\alpha=e^2/(4\pi\varepsilon_0 c)
\simeq1/137$ is the fine-structure constant.
Recall that for graphene $v_{\rm F}/c\sim 1/300$, and the effective
fine-structure constant in eq. (\ref{2.24}) is $\alpha_{\rm
eff}=\alpha\frac{c}{v_{\rm F}} \sim 2 $. Thus, the honeycomb lattice
provides us with an interesting strong-coupling theory.

It is worth noting that in the case of finite temperature, and/or
finite density, polarization effects and a Debye screening mass may
considerably modify the bare Coulomb potential \cite{n19,10} leading
to a full (non-perturbative) expression $U^C(x)$.

It should be noted that a great simplification in solving the
Hartree-Fock (gap) equation for fermion masses and the
Bethe-Salpeter equation for exciton bound states arises, if one
approximately replaces the unknown full Coulomb potential $U^C(x)$
by a $\delta$-function contact interaction. In particular, let us
suppose that the photon propagator gets a non-perturbative
effective photon mass $M$. \footnote{ Photon masses might, for example, arise from a Higgs mechanism with a Cooper pair condensate (Meissner effect) and/or from Debye screening. In the following, we shall consider $M$ rather as the size of the low-energy region, where a contact
approximation is applicable. } In this case, the integral in eq.
(\ref{2.24}) is replaced by
\begin{equation}
\int{\frac{d^2k}{(2\pi)^2}}\exp(i\vec k\cdot\vec
r)\frac{1}{\sqrt{{\vec k}^2+M^2}}\GoesToWhen\limits_{M^2\gg \vec
  k^2}\frac{1}{M}\delta^{(2)}(\vec r)
.\label{2.25}\end{equation} Thus we get a "low-momentum" contact
interaction
\begin{equation}
U^C(x)=\frac{2\pi\alpha}{M}\frac{c}{v_{\rm
F}}\delta^{(3)}(x)\equiv G_c\delta^{(3)}(x)
,\label{2.26}\end{equation} with
$G_c=\frac{2\pi\alpha}{M}\frac{c}{v_{\rm F}}$ characterizing the
effective interaction strength.
\begin{figure}
\begin{center}
\includegraphics[width=200pt]{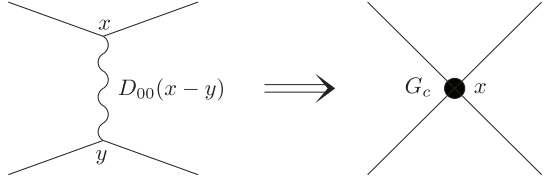}
\caption{\it{Contact approximation to the non-local Coulomb
    interaction. The full photon propagator $D_{00}$ is replaced by a
    local four-fermion interaction of strength $G_c$.}}
\label{Fig.2}
\end{center}
\end{figure}
The considered contact approximation (see Fig. \ref{Fig.2}) leads to
the following $U(2N_{\rm f})$-invariant four-fermion part of the
interaction Lagrangian \footnote{ Clearly, due to suppression of
spatial components of currents and associated retardation effects,
the expression (\ref{2.27}) is not Lorentz-invariant. It thus
differs from the Thirring model with $v_{\rm
  F}=c$ considered in refs. \cite{11,n21}.}
\begin{equation}
L^C_{\rm int}=-\frac{G_c v_{\rm F}}{2}\left[\psibar(x)\gamma^0\psi(x)\right]^2 .\label{2.27}
\end{equation}
Since the coupling constant $G_c$ has mass dimension
$\left[G_c\right]=-1$, a theory based on eq. (\ref{2.27}) is not
renormalizable in usual perturbation theory, but turns out to be
renormalizable in the $1/N_{\rm f}$ expansion.

In the following subsection we shall project the 4-fermion
Coulomb-based interaction (\ref{2.27}), motivated from photon
exchange, by a Fierz transformation into fermion-hole channels,
where bound exciton states occur. This results in an effective
Nambu--Jona-Lasinio (NJL) type of interaction which keeps global
chiral $U(2N_{\rm f})$ and fermion number $U(1)$ symmetries intact. It is
well-known from strong interaction physics that NJL-type models
\cite{n6} naturally incorporate the dynamical mechanism for
spontaneous breakdown and restoration of chiral symmetry. As
demonstrated for the case of QCD, an analogous contact approximation
for the long-range gluon interaction between colored quark currents
leads to a QCD-based NJL model which has been shown to give a
successful description of fermion masses and coupling constants, as
well as of masses of hadronic bound states \cite{n9,15}. Based on
such experience, one might expect that the contact interaction
(\ref{2.27}) and its resulting Fierz-transformed NJL-type terms
might become, at least qualitatively, a reasonable starting point
for a non-perturbative low-energy description of chiral symmetry
breaking and restoration in graphene-like models.

It should be noted that the $U(2N_{\rm f})$ symmetry of the Lagrangian is
broken, when a spin Zeeman interaction with an external magnetic
field $\bf B$ is included (comp. Section 2.3.3). Moreover, it is
worth mentioning that the valley-sublattice symmetry $U(2)_{vs}$ is
an emergent symmetry arising in the continuum limit, which is not
exact in the tight-binding lattice Hamiltonian $H_0$ \cite{n24,n25}.
By this reason, one cannot postulate that chiral $U(2N_{\rm f})$
invariance of the continuum Lagrangian $L$, corresponding to the
action (\ref{2.23}) or the contact Coulomb-like interaction
(\ref{2.27}), must hold for the complete effective Lagrangian, too.

In particular, the Coulomb interaction on the lattice contains
additionally a small on-site repulsive interaction term
\cite{n19,n26}
\begin{equation}
\Delta L_{\rm int}=\frac{Gv_{\rm F}}{2}(\psibar\psi)^2
,\label{2.28}\end{equation} which breaks the $U(2N_{\rm f})$ symmetry
explicitly according to $U(2N_{\rm f})\to U(N_{\rm f})_{t^0}\otimes
U(N_{\rm f})_{t^3}$. Here the groups $U(N_{\rm f})_{t^i}$ have the Lie algebra of eq. (\ref{2.17}) with two unbroken generators $t^0$, $t^3$.

Analogous terms may arise from phonon-mediated interactions with
coupling strength $g$ \cite{n33}. Combining the expressions
(\ref{2.27}) and (\ref{2.28})
and including the phonon-mediated interaction term
leads to the symmetry breaking
interaction Lagrangian
\begin{equation}
L_{\rm int}=-\frac{1}{2}G_c v_{\rm F}(\psibar \gamma^0\psi)^2+
\frac{\widetilde{G} v_{\rm F}}{2}(\psibar\psi)^2
,\label{2.29}\end{equation} where $\widetilde{G}$ is the effective
coupling $\widetilde{G}=G+g$.

\subsubsection{Fierz-transformed interaction Lagrangian}\label{fierz}

In the following we shall study dynamically generated fermion masses
(gaps) arising from condensates of exciton fields describing
quasiparticle-hole bound states. This requires to project the
interaction term (\ref{2.29}) by a Fierz transformation into
bound-state channels for exciton fields $\varphi^A$,
$\varphi^A\sim\psibar\Gamma^A\psi$. Here $\Gamma_A$ is a complete
basis of the $4\times 4$ Dirac algebra, given by the 16 matrices
\begin{equation}
\left\{\Gamma^A\right\}_{A=1}^{16}=\left\{{\rm
I}_4,i\gamma^3,\gamma^5,\gamma^{35},\widetilde{\gamma}^{\mu},
\widetilde{\gamma}^{\mu 3},\widetilde{\gamma}^{\mu
5},\widetilde{\gamma}^{\mu 35}\right\} , \label{2.30}\end{equation}
where
\begin{eqnarray}
\widetilde{\gamma}^{\mu}=(\gamma^0,i\gamma^k),~~~&&
\widetilde{\gamma}^{\mu 3}=(\gamma^0\gamma^3,i\gamma^k\gamma^3),\nonumber\\
\widetilde{\gamma}^{\mu 5}=(\gamma^0
i\gamma^5,\gamma^k\gamma^5),~~~&& \widetilde{\gamma}^{\mu
35}=(\gamma^0\gamma^{35},i\gamma^k\gamma^{35})
,~~~(k=1,2)\label{2.31}
\end{eqnarray}
and we have
$\left(\Gamma^A\right)^+=\Gamma^A=\left(\Gamma^A\right)^{-1}$, ${\rm
Tr}\,\Gamma^A\Gamma^B=4\delta^{AB}$.

Taking into account only NJL-type scalar/pseudoscalar interactions
and discarding, for simplicity, axial/vector type terms, the Fierz
transformation of the two terms in the interaction Lagrangian
(\ref{2.29}) gives (see Appendix A, eqs. (\ref{A.9}) and (\ref{A.10})):
\begin{equation}
L^{F,C}_{\rm int}=\frac{1}{2}\frac{G_c v_{\rm F}}{4N_{\rm f}}\left\{\left[
(\psibar\psi)^2+(\psibar\gamma^3\psi)^2+(\psibar
i\gamma^5\psi)^2\right] +(\psibar\gamma^{35}\psi)^2\right\}+... ,
\label{2.32}\end{equation}
\begin{equation}
\Delta L^{F}_{\rm int}=-\frac{1}{2}\frac{\widetilde{G} v_{\rm
F}}{4N_{\rm f}} \left\{\left[
(\psibar\psi)^2-(\psibar\gamma^3\psi)^2-(\psibar
i\gamma^5\psi)^2\right]+(\psibar\gamma^{35}\psi)^2 \right\}+... .
\label{2.33}\end{equation}

Moreover, since we shall consider only the condensates of
flavor/spin-singlet excitons, we have taken into account only
corresponding singlet terms in the completeness relations
(\ref{A.4}) and (\ref{A.5}).

Notice that the first three terms in the $U(2N_{\rm f})$ invariant Coulomb
contact interaction (\ref{2.32}) are just the scalar product of the
$U(2)_{vs}$-vector $\vec V_i=(\psibar \Gamma_i \psi)$,
$\Gamma_i=\left\{{\rm I}_4,\gamma^3,i\gamma^5\right\}$, whereas
$\psibar\gamma^{35}\psi$ is a scalar. Taking into account eqs.
(\ref{2.29}), (\ref{2.32}) and (\ref{2.33}) leads to the following
effective low-energy four-fermion Lagrangian
\begin{equation}
L=\psibar i\lefteqn{/}\widetilde{\partial}\psi+ \frac{G'_1 v_{\rm
F}}{2N_{\rm f}} \left[ (\psibar\psi)^2+(\psibar\gamma^{35}\psi)^2 \right]+
\frac{G'_2 v_{\rm F}}{2N_{\rm f}} \left[ (\psibar\gamma^3\psi)^2+(\psibar
i\gamma^{5}\psi)^2 \right] ,\label{2.34}\end{equation} where
$\lefteqn{/}\widetilde{\partial}= \gamma^\mu\tilde\partial_\mu$, and
$G'_1=\frac{1}{4}(G_c-\widetilde{G})$,
$G'_2=\frac{1}{4}(G_c+\widetilde{G})$.

Clearly, the above approximation scheme does not allow a
determination of the effective coupling constants $G_1'$, $G_2'$
from an underlying microscopic lattice theory, eventually including
lattice vibrations. By this reason, the above four-fermion model
(\ref{2.34}) can be only considered as a schematic one. For the
following general considerations of possible exciton condensates and
related phase transitions, it turns out to be reasonable to
generalize this model further by omitting from now on any symmetry
constraints between coupling constants. This leads us to the
following schematic low-energy model for interacting fermions on a
hexagonal lattice
\begin{eqnarray}
L&=&L_0+L_{int}=\bar \psi
i\lefteqn{/}\widetilde{\partial}\psi\nonumber\\
&+&\Big\{\frac{1}{2N_{\rm f}}G_1v_{\rm F}(\bar\psi\psi)^2+\frac{1}{2N_{\rm f}}G_2v_{\rm F}(\bar\psi\gamma^{35}\psi)^2+\frac{1}{2N_{\rm f}}H_1v_{\rm F}(\bar\psi
i\gamma^5\psi)^2+\frac{1}{2N_{\rm f}}H_2v_{\rm F}(\bar\psi\gamma^3\psi)^2\Big\}.
\label{2.35}
\end{eqnarray}
Note once more that $\psi (x)$ in (\ref{2.35}) transforms as a
fundamental multiplet of the flavor $U(N_{\rm f})$ group, i.e. $\psi
(x)\equiv\psi^a(x)$, where $a=1,...,N_{\rm f}$. Moreover, each component
of this multiplet is a four-component Dirac spinor. (Both, the
trivial summation over flavor ($a=1,...,N_{\rm f}$) and the summation over
spinor indices in (\ref{2.35}) are implied.) An extended
four-fermion model of this type was studied in papers \cite{n13,n27}
for (2+1)D QED and QCD in external magnetic and chromomagnetic
fields. However, these papers have no direct physical relation to
the considered honeycomb tight-binding model, its low-energy
expansion, the Fierz-transformed contact Coulomb and phonon
interactions and also do not use the naturally arising chiral (Weyl)
representation of the Dirac algebra. Nevertheless, we can use their
methods and results to our investigation of the effective potential,
the solution of gap equations and the exciton mass spectrum (see
Section 3.1).

It is worth noting that the general Lagrangian (\ref{2.35}) is
invariant under spatial inversion $\mathcal P$ (see eq.
(\ref{2.18})) and discrete chiral transformations
\begin{eqnarray}
\gamma^5~:~~\psi\to\gamma^5\psi,~~\bar\psi\to-\bar\psi\gamma^5;~~~~~~
\gamma^3~:~~\psi\to\gamma^3\psi,~~\bar\psi\to\bar\psi\gamma^3.
\label{2.36}
\end{eqnarray}
For completeness and later applications, we finally introduce
external magnetic fields and the chemical potential into the
Lagrangian (\ref{2.35}).

\subsubsection{External magnetic fields}\label{magnit}

Let us apply the following substitutions
$\partial_0\to(\partial_0-i\mu)$, $\partial_k\to(\partial_k+ieA_k)$
($k=1,2$) to the kinetic part $L_0$ in eq. (\ref{2.35}), i.e.
\footnote{The chemical potential $\mu$ arises in the tight-binding
lattice model in a natural way, if one includes in eq. (\ref{2.1}) the
next-to-nearest (in-sublattice) hoppings $\Delta
H_0=-t'\sum\limits_{\ll i,j \gg}(\psi^{+A}(\vec r_i)\psi^{A}(\vec
r_j)+\psi^{+B}(\vec r_i)\psi^{B}(\vec r_j))$, leading to $\mu=3t'$.
Clearly, such term violates the quasiparticle-hole symmetry. }
\begin{eqnarray}
\bar \psi i
\tilde{/{\hspace{-1.8mm}\partial}}\psi&\to&\bar\psi\Big
[i\gamma^0(\partial_0-i\mu+i\frac{g}{2}\mu_B\vec\sigma\cdot\vec
B)+i\gamma^1v_{\rm F}(\partial_x+ieA_x)+i\gamma^2v_{\rm F}(\partial_y+ieA_y)\Big
]\psi,\label{2.37}
\end{eqnarray}
where $A_x$ and $A_y$ are components of the external electromagnetic
vector potential. The additional term $\sim\vec\sigma\cdot\vec B$ in
eq. (\ref{2.37}) is the Zeeman energy term, which describes the
(nonrelativistic) interaction of the real spin of quasiparticles
with the magnetic field $\vec B$ and has to be added separately.
Here $g$ is the spectroscopic Land\'e factor, and $\mu_B=e/(2m)$ is
the Bohr magneton.

Let us now consider a tilted magnetic field $\vec
B=(B_\parallel,0,B_\perp)$ with in-plane component $B_\parallel$ in
$x$-direction and transversal component $B_\perp$ in the
$z$-direction, transversal to the plane. (Phase transitions in
planar systems under the influence of a tilted magnetic field and
Zeeman interaction were recently considered in eq. \cite{n12}.) It is
convenient to choose a gauge, where the three-dimensional vector
potential takes the form $\vec A=(0,{\cal A}_2+B_\perp x,B_\parallel
y)$ with constant ${\cal A}_2$, so that indeed $\vec B=\rot \vec A$.
Obviously, the transversal component $B_\perp$ couples with the
orbital angular momentum $L_z$ and spin component $\frac
12\sigma_z$, whereas the parallel component $B_\parallel$ couples
only to the spin of quasiparticles. Note that we admitted also a
constant gauge field component ${\cal A}_2$ for later use, when we
shall compactify the $y$-coordinate to get a nanotube cylinder. In
such a case, a constant field component in the covariant derivative
of the compactified direction cannot be gauged away. In particular,
${\cal A}_2$ turns out to play an important role for the description
of the Aharonov-Bohm (AB) effect in hexagonal lattice nanotubes (see
Sect. 4).

As is well known, the transversal field $B_\perp$ leads to Landau
levels of fermions and the very interesting quantum Hall effect (for
details, see, e.g., ref. \cite{21}). In the present paper, we shall,
however, take $B_\perp =0$ and discuss the fermionic gap equations
and exciton masses in the general schematic model (\ref{2.35}).
Moreover, for illustrations and nontrivial application, we shall
consider later on the simple Gross-Neveu (GN) version of the
Lagrangian (\ref{2.35}) with $G_1\ne 0$, $G_2=H_1=H_2=0$, and
investigate phase transitions in nanotubes in dependence on the
AB-field ${\cal A}_2$, the Zeeman interaction with $\vec B$ chosen
parallel to the cylinder axis and taking a finite chemical potential
$\mu$ and temperature $T$ (see Sect. 4). \footnote{Phase transitions
in  (2+1)-dimensional GN-models including magnetic fields have also
been studied in numerous earlier papers \cite{n11,24,250,251,252}.}

\subsection{Exciton fields, gap equations and symmetry breaking}\label{exiton}

Let us now rewrite the Lagrangian (\ref{2.35}) by introducing
(auxiliary) excitonic fields $\sigma_1$, $\sigma_2$, $\varphi_1$,
and $\varphi_2$ via the Hubbard-Stratonovich transformation
\begin{eqnarray}
{\cal L}[\bar\psi,\psi,\sigma_i,\varphi_i]&=&\bar \psi\Big [i\lefteqn{/}\widetilde{\partial} -\sigma_1-\sigma_2\gamma^{35}-\varphi_1i\gamma^5-\varphi_2\gamma^3\Big ]\psi-
N_{\rm f}\sum_{k=1}^2\left
(\frac{\sigma_k^2}{4v_{\rm F}G_k}+\frac{\varphi_k^2}{4v_{\rm F}H_k}\right ).
\label{2.38}
\end{eqnarray}
Obviously, by inserting the field equations for excitonic fields
\begin{eqnarray}
\sigma_1=-2\frac{G_1v_{\rm F}}{N_{\rm f}}\bar\psi\psi,~\sigma_2=-2\frac{G_2v_{\rm F}}{N_{\rm f}}\bar\psi\gamma^{35}\psi,~\varphi_1=-2\frac{H_1v_{\rm F}}{N_{\rm f}}\bar\psi
i\gamma^5\psi,~\varphi_2=-2\frac{H_2v_{\rm F}}{N_{\rm f}}\bar\psi\gamma^3\psi
\label{2.39}
\end{eqnarray}
back into expression (\ref{2.38}), we reproduce the Lagrangian
(\ref{2.35}).

In order to get the Hartree-Fock gap equations for dynamical fermion
masses in terms of exciton or fermion condensates, we shall take the
vacuum (ground state) expectation values $\langle\cdots\rangle$ on
both sides of the expressions in eq. (\ref{2.39}). This leads us to
the gap equations
\begin{eqnarray}
\label{2.40a}\langle\sigma_1\rangle&=&-2\frac{G_1v_{\rm F}}{N_{\rm f}}\langle\bar\psi\psi\rangle=2\frac{G_1v_{\rm F}}{N_{\rm f}}{\rm Tr}_{sf}\left [iG(x,x)\right ],\\ \label{2.40b}\langle\sigma_2\rangle &=&-2\frac{G_2v_{\rm F}}{N_{\rm f}}\langle\bar\psi\gamma^{35}\psi\rangle =2\frac{G_2v_{\rm F}}{N_{\rm f}}{\rm Tr}_{sf}\left [\gamma^{35}iG(x,x)\right ],\\
\label{2.40c}\langle\varphi_1\rangle
&=&-2\frac{H_1v_{\rm F}}{N_{\rm f}}\langle\bar\psi i\gamma^5\psi\rangle
=2\frac{H_1v_{\rm F}}{N_{\rm f}}{\rm Tr}_{sf}\left [i\gamma^{5}iG(x,x)\right ],\\
\label{2.40d}\langle\varphi_2\rangle
&=&-2\frac{H_2v_{\rm F}}{N_{\rm f}}\langle\bar\psi \gamma^3\psi\rangle
=2\frac{H_2v_{\rm F}}{N_{\rm f}}{\rm Tr}_{sf}\left [\gamma^{3}iG(x,x)\right ],
\end{eqnarray}
where $G(x,y)$ ($x=(x^0,\vec r)$) is the full quasiparticle
propagator which is proportional to the unit matrix in the
$N_{\rm f}$-dimensional flavor space as well as a 4$\times$4 matrix acting
in the 4-dimensional spinor space. So the symbol ${\rm Tr}_{sf}$ in
eqs. (\ref{2.40a})-(\ref{2.40d}) means the trace of an operator just
over the spinor (s) and flavor (f) spaces. The inverse propagator
with excitonic condensates $G^{-1}(x,x')$
 has the following matrix elements
in the direct product of the flavor ($a,b=1,...,N_{\rm f}$) and spinor
spaces ($\alpha ,\beta=1,...,4$)
\begin{eqnarray}
\big [G^{-1}(x,x')\big ]_{\alpha\beta}^{ab}&=&\Big
[i\lefteqn{/}\widetilde{\partial}
 -\langle\sigma_1\rangle-\langle\sigma_2\rangle\gamma^{35}-\langle\varphi_1\rangle i\gamma^5-\langle\varphi_2\rangle\gamma^3\Big ]_{\alpha\beta}\delta^{ab}\delta^{(3)}(x-x'),
\label{2.41}
\end{eqnarray}
where $i\lefteqn{/}\widetilde{\partial}=\gamma^0 i\partial_0+i
v_{\rm F}\vec\gamma\cdot\vec\nabla$. It is clear that each trace over the
flavor space in eqs. (\ref{2.40a})-(\ref{2.40d}) gives there the
factor $N_{\rm f}$. Due to this reason, the solutions of the gap equations
(\ref{2.40a})-(\ref{2.40d}), i.e. the condensates
$\langle\sigma_1\rangle$, $\langle\sigma_2\rangle$ etc., do not
depend on $N_{\rm f}$.

Let us finally quote the symmetry breaking properties of the
condensates concerning chiral $U(2N_{\rm f})$ transformations and discrete
$\mathcal P$, $\mathcal C$, $\mathcal T$, $\gamma^5$ and $\gamma^3$
transformations (see Table \ref{Tab1}) \cite{n5}:

(i)  $\langle\bar\psi\psi\rangle$ -- it breaks $U(2N_{\rm f})$ and
discrete $\gamma^5$, $\gamma^3$ transformations, but preserves
$\mathcal P$, $\mathcal C$, $\mathcal T$.

(ii)  $\langle\bar\psi\gamma^{35}\psi\rangle$ -- it preserves
$U(2N_{\rm f})$ and $\mathcal C$, $\gamma^5$ and $\gamma^3$, but breaks
$\mathcal P$ and $\mathcal T$. The related ``Haldane mass''
$m_2=\langle\sigma_2\rangle /v_{\rm F}^2$ is related to the parity anomaly in (2+1) dimensions \cite{26} (see, also ref. \cite{Charlier}).

(iii) $\langle\bar\psi i\gamma^5\psi\rangle$ -- it breaks $U(2N_{\rm f})$
and discrete $\mathcal P$, $\mathcal C$, $\gamma^5$, but preserves
$\mathcal T$ and $\gamma^3$.

(iv) $\langle\bar\psi \gamma^3\psi\rangle$ -- it breaks $U(2N_{\rm f})$
and $\gamma^3$, but preserves $\mathcal P$, $\mathcal C$, $\mathcal
T$ and $\gamma^5$.\\
Below we demonstrate (see Section 3.1.1 and,  especially, Appendix B)
that, depending on the values of the coupling constants, five
different phases may be implemented in the framework of the model (\ref{2.35}).
 One of them is a trivial one, because its ground state is
characterized by zero values of all condensates $\langle\sigma_{1,2}
\rangle$ and $\langle\varphi_{1,2}\rangle$ and, therefore, has a
highest possible symmetry. In each of the remaining phases only one
of these condensates has a nonzero value. Hence, the ground states
of these nontrivial phases of the model (\ref{2.35}) differ in their
symmetry properties (see  Table \ref{Tab1} and/or the above points
(i),..., (iv)).

Before concluding this Section, it is worth mentioning that the
axial/vector interactions appearing in a general Fierz-transformed
interaction term (compare eqs. (\ref{A.9})-(\ref{A.11})) may
generate additional chemical potentials dynamically. Applying again
the Hubbard-Stratonovich transformation and introducing axial/vector
exciton fields $a_\mu$, one gets an additional term
\begin{eqnarray}
\Delta{\cal L}=\sum_{k=0}^3a_\mu^k\bar\psi\gamma^\mu
t^k\psi-N_{\rm f}\sum_{k=0}^3\frac{1}{4v_{\rm F}\widetilde G_k}(a_\mu^k)^2
\label{2.42}
\end{eqnarray}
with general coupling constants $\widetilde G_k$. Admitting
nonvanishing condensates $\langle a_0^k\rangle$, which have to be
determined by respective gap equations, then provides additional
dynamical chemical potentials $\mu^k=i\langle a_0^k\rangle$. The
investigation of the combined system of gap equations of masses and
chemical potentials is planned elsewhere.
\renewcommand{\arraystretch}{0.9}
\renewcommand{\tabcolsep}{0.5cm}
\begin{table*}[!t]
\centering
\begin{tabular}{|c | c c c c|}\hline\hline
$\langle\bar\psi\Gamma_i\psi\rangle$      &
$\langle\bar\psi\psi\rangle$   &
$\langle\bar\psi\gamma^{35}\psi\rangle$   &   $\langle\bar\psi
i\gamma^5\psi\rangle$   &  $\langle\bar\psi \gamma^3\psi\rangle$
\\ \hline
$\mathcal P$   &    1   &   -1   &   -1   &   1  \\
$\mathcal C$  &   1   &  1    &   -1   &  1  \\
$\mathcal T$  & 1  &  -1   &   1   &  1  \\ \hline
$\gamma^5$  & -1   & 1   &   -1   &  1  \\
$\gamma^3$  &  -1  & 1   & 1   &   -1   \\
 \hline\hline
\end{tabular}
\caption{\it\label{Tab1} Transformation properties of various
condensates
  $\langle\bar\psi\Gamma_i\psi\rangle$, where
now $\Gamma_i=\{ {\rm I}_4,\gamma^{35},i\gamma^5, \gamma^3$\}, under
discrete $\mathcal P$, $\mathcal C$, $\mathcal T$ and $\gamma^5$,
$\gamma^3$ transformations (here we consider ${\mathcal
P}:~(x^0,x,y)\to (x^0,-x,y)$).}
\end{table*}

\section{Effective potential: general definitions \label{sec3}}
\subsection{Hexagonal lattice sheets}

Let us consider the partition function of the semi-bosonized
Lagrangian (\ref{2.38}) given by the path integral
\begin{eqnarray}
Z=\int D\bar\psi D\psi\int D\sigma_1 D\sigma_2 D\varphi_1 D\varphi_2
\exp\left \{i\int dx^0d^2x {\cal
L}[\bar\psi,\psi,\sigma_i,\varphi_i]\right\}. \label{3.1}
\end{eqnarray}
Integrating in eq. (\ref{3.1}) over fermion fields
and rewriting the resulting determinant of the Dirac operator
$\hat D(x,y)=D(x,y)I_{N_f}$
(being the inverse propagator given by eq. (2.46))
as $\Det(\hat D)=(\Det D)^{N_f}=\exp (N_f{\rm Tr}_{sx}\ln D)$,
one obtains
\begin{eqnarray}
Z&=&\int D\sigma_1 D\sigma_2 D\varphi_1 D\varphi_2 \exp\left \{iN_{\rm f} S_{\rm eff}(\sigma_i,\varphi_i)\right\},\nonumber \\
S_{\rm eff}(\sigma_i,\varphi_i)&=&-\int dx^0d^2x\sum_{k=1}^2\left
(\frac{\sigma_k^2}{4v_{\rm F}G_k}+\frac{\varphi_k^2}{4v_{\rm F}H_k}\right
)\nonumber \\&-& i {\rm Tr}_{sx} \ln
(i\lefteqn{/}\widetilde{\partial}
-\sigma_1-\sigma_2\gamma^{35}-\varphi_1i\gamma^5-\varphi_2\gamma^3
).
\label{3.2}
\end{eqnarray}
Here the  quantity $S_{\rm eff}(\sigma_i,\varphi_i)$ (\ref{3.2}) is
the effective action of the model and the Tr$_{sx}$-operation stands
for the trace in four-dimensional spinor (s) and (2+1)-dimensional
coordinate (x) spaces, respectively. Note that the expression
(\ref{3.2}) for $S_{\rm eff}$ can be used in order to generate
one-particle irreducible Green functions of the exciton fields
$\sigma_i(x)$ and $\varphi_i(x)$ in the leading order of the
large-$N_{\rm f}$ expansion \cite{28}. Moreover, the effective
potential of the model is obtained by taking the effective action
$S_{\rm eff}$ in the path integral (\ref{3.2}) at the saddle point
${\sigma_i,\varphi_i={\rm const}}$,
\begin{eqnarray}
\Veff(\sigma_i,\varphi_i)\int
dx^0d^2x&=&-S_{\rm eff}(\sigma_i,\varphi_i)\Big
|_{\sigma_i,\varphi_i={\rm const}}. \label{3.4}
\end{eqnarray}
Following the technique of e.g. the paper \cite{ek},
it is possible
to find from eqs. (\ref{3.2}) and (\ref{3.4}) that
\begin{eqnarray}
\Veff(\sigma_i,\varphi_i) &=&\sum_{k=1}^2\left
(\frac{\sigma_k^2}{4v_{\rm F}G_k}+\frac{\varphi_k^2}{4v_{\rm F}H_k}\right
)+i\int\frac{dp_0d^2\vec p}{(2\pi)^3}{\rm Tr}_{s} \ln D(p),
\label{3.40}
\end{eqnarray}
where $D(p)=p_0\gamma^0-v_{\rm F}\vec p\vec\gamma
-\sigma_1-\sigma_2\gamma^{35}-\varphi_1i\gamma^5-\varphi_2\gamma^3$
is the Fourier transformation of the flavor-independent part $D(x,y)$ of the above Dirac operator. Since ${\rm Tr}_{s} \ln
D(p)=\ln {\rm Det}D(p)=\sum_i\ln\epsilon_i$, where $\epsilon_i$ are the four eigenvalues of the 4$\times$4 matrix $D(p)$,
\begin{eqnarray}
\epsilon_{1,2,3,4}=\sigma_1\pm\sqrt{\left
(\sigma_2\pm\sqrt{p^2_0-v_{\rm F}^2\vec p^2}\right
)^2-\varphi_1^2-\varphi_2^2}~, \label{3.41}
\end{eqnarray}
we have from eq. (\ref{3.40})
\begin{eqnarray}
\Veff(\sigma_i,\varphi_i)
&=&\sum_{k=1}^2\left\{\frac{\sigma_k^2}{4v_{\rm F}G_k}+\frac{\varphi_k^2}{4v_{\rm F}H_k}+i\int\frac{dp_0d^2\vec
p}{(2\pi)^3}\ln\left (p_0^2-v_{\rm F}^2\vec p^2-M_k^2\right )\right\},
\label{3.5}
\end{eqnarray}
where $M_{1,2}=|\sigma_2\pm\rho|$,
$\rho=\sqrt{\sigma_1^2+\varphi_1^2+\varphi_2^2}$. Integration over
$p_0$ in eq. (\ref{3.5}) can now be performed by using the general
relation $\int dp_0\ln(p_0-A)=i\pi|A|$, which is true up to an
infinite term independent of the real quantity $A$. So we have from
eq. (\ref{3.5})
\begin{eqnarray}
\Veff(\sigma_i,\varphi_i)
&=&\sum_{k=1}^2\left\{\frac{\sigma_k^2}{4v_{\rm F}G_k}+\frac{\varphi_k^2}{4v_{\rm F}H_k}
-\int\frac{d^2\vec p}{(2\pi)^2} \sqrt{v_{\rm F}^2\vec p^2+M_k^2}\right\}.
\label{3.51}
\end{eqnarray}
Since the integral term in this formula is an ultraviolet divergent
improper integral, the effective potential
$\Veff(\sigma_i,\varphi_i)$ is an ultraviolet divergent quantity.
One way to obtain from eq. (\ref{3.51}) a finite expression for the
effective potential is to regularize it by simply integrating in eq.
(\ref{3.51}) over the cutted region, $|\vec p|<\Lambda$, in polar
coordinates, where the cutoff parameter $\Lambda={\cal O}(1/a)$ is
of the order of the inverse lattice spacing. As a result, we have
for the regularized effective potential
\begin{eqnarray}
\Veff(\sigma_i,\varphi_i)
&=&\sum_{k=1}^2\left\{\frac{\sigma_k^2}{4v_{\rm F}}\left
    (\frac{1}{G_k}-\frac{2\Lambda}{\pi}\right
  )+\frac{\varphi_k^2}{4v_{\rm F}}\left
    (\frac{1}{H_k}-\frac{2\Lambda}{\pi}\right )+\frac{M_k^3}{6\pi
    v^2_{\rm F}}+M_k^3/v_{\rm F}^3~{\cal O}\left (\frac{M_k}{\Lambda}\right
  )\right\}.
\label{3.7}
\end{eqnarray}
In eq. (\ref{3.7}) we have omitted constant terms, which do not
depend on the dynamical parameters $M_k$. It is well known that
coordinates of the global minimum point of the effective potential
supply us with condensates $\langle\sigma_1\rangle$, etc., as well
as with a phase structure of the model. In order to simplify the
investigation of the function (\ref{3.7}) on the global minimum
point, we now suppose that $M_k/\Lambda\ll 1$. In this case the last
term in eq. (\ref{3.7}) can also be omitted and
$\Veff(\sigma_i,\varphi_i)$ takes the form
\begin{eqnarray}
\Veff(\sigma_i,\varphi_i)
&=&\sum_{k=1}^2\left\{\frac{g_k\sigma_k^2}{4v_{\rm F}}+\frac{h_k\varphi_k^2}{4v_{\rm F}}+\frac{M_k^3}{6\pi
v^2_{\rm F}}\right\}, \label{3.9}
\end{eqnarray}
where we have used the notations
\begin{eqnarray}
g_k=\frac{1}{G_k}-\frac{1}{G_c},~~h_k=\frac{1}{H_k}-\frac{1}{H_c}
\label{3.10}
\end{eqnarray}
and $G_c^{-1}=H_c^{-1}=\frac{2\Lambda}{\pi}$. Assuming that $G_k$,
$H_k$ and $\Lambda$ are effective finite quantities, with $\Lambda$
restricting the low-energy region of applicability, one can use for
the effective potential just the finite expression (\ref{3.9}). As a
result, we see that in this case both the phase structure of the
model and the condensates $\langle\sigma_1\rangle$, etc. are
described (instead of the bare quantities $G_k$, $H_k$ and the
cutoff $\Lambda$) in terms of the finite quantities $g_k$ and $h_k$.

There is yet another way to get a finite effective potential from
the formally divergent expression (\ref{3.51}). It is based on the
fact that (2+1)-dimensional quantum field theories with four-fermion
interactions are renormalizable in the framework of the large-$N_{\rm f}$
expansion technique \cite{28}. So, following, e.g., the procedure of
refs. \cite{n13,n27}, one can renormalize the quantity (\ref{3.51})
by introducing the cutoff parameter $\Lambda$, the renormalization
scale $m$ and renormalized coupling constants $g_k(m)$, $h_k(m)$. It
turns out that in this case the finite renormalized expression for
$\Veff(\sigma_i,\varphi_i)$ looks like the effective potential in
eqs. (\ref{3.9})-(\ref{3.10}), in which the relations
\begin{eqnarray}
g_k=\frac{1}{g_k(m)}-\frac{2m}{\pi},~~h_k=\frac{1}{h_k(m)}-\frac{2m}{\pi}
\label{3.101}
\end{eqnarray}
are valid also in addition to eq. (\ref{3.10}). It is clear from
eqs. (\ref{3.10})-(\ref{3.101}) that in this case the parameters
$g_k$ and $h_k$ are independent of both  the renormalization scale
$m$ and the cutoff parameter $\Lambda$, i.e. they are finite and
renormalization invariant parameters.  As a result, the obtained
renormalized effective potential (\ref{3.9}) is also a
renormalization group invariant quantity.

\subsubsection{Gap equations and fermion masses}

It is clear from the expression (\ref{2.41}) for the fermion
quasiparticle propagator that there might exist several dynamical
fermion masses,
$m_i=\langle\sigma_i\rangle /v_{\rm F}^2$ or
$m'_i=\langle\varphi_i\rangle/v_{\rm F}^2$.
 (The relation between dynamical masses and
energy gap in graphene-like condensed-matter systems and nanotubes
is discussed, e.g. in ref. \cite{Charlier}.) Since the excitonic
condensates $\langle\sigma_i\rangle$ and $\langle\varphi_i\rangle$
are determined by the global minimum point
$(\sigma^0_i,\varphi^0_i)$ of the effective potential (\ref{3.9}),
it is necessary to study its stationarity (gap) equations
\begin{eqnarray}
\frac{\partial
\Veff(\sigma_i,\varphi_i)}{\partial\sigma_i}=0,~~~\frac{\partial
\Veff(\sigma_i,\varphi_i)}{\partial\varphi_i}=0,~~i=1,2
\label{3.11}
\end{eqnarray}
in order to find dynamical masses $m_i$, $m'_i$ of fermion
quasiparticles. (Note that in general the system of the gap
equations (\ref{2.40a})-(\ref{2.40d}) is equivalent to the
stationarity equations (\ref{3.11}).) The global minimum point
$(\sigma^0_i,\varphi^0_i)$ of the effective potential (\ref{3.9})
determines just the condensate values, i.e.
$\sigma^0_i=\langle\sigma_i\rangle$,
$\varphi^0_i=\langle\varphi_i\rangle$. (For a detailed discussion of
possible condensates, dynamical fermion masses and related phase
structure in dependence on general choices of coupling constants
$g_i$, $h_i$ we refer to refs. \cite{n13,n27} and also to Appendix B.
There the extremum properties of the effective potential (\ref{3.9})
are investigated in the most general case, i.e. for arbitrary
relations between coupling constants).

For illustrations, let us specify for a moment to the $U(2)\times
U(N_{\rm f})$-symmetric model with $g_1=g_2=h_1=h_2\equiv g$. As is easily
seen from eq. (\ref{3.10}), in this particular case we have
$G_i=H_i\equiv G$ ($i=1,2$). Moreover, in this symmetric case
$\Veff$ (\ref{3.9}) simplifies to
\begin{eqnarray}
\Veff(\sigma_i,\varphi_i)&=&\frac{g}{4v_{\rm F}}\cdot\frac{M_1^2+M_2^2}{2}+\frac{M_1^3}{6\pi
v^2_{\rm F}}+\frac{M_2^3}{6\pi v^2_{\rm F}}~, \label{3.12}
\end{eqnarray}
where $M_{1,2}$ are given just after the expression (\ref{3.5}). In
addition, it is clear from eq. (\ref{3.12}) that in this specific
case the effective potential depends on the $O(3)$ - invariant
$\rho=\sqrt{\sigma_1^2+\varphi_1^2+\varphi_2^2}$. So, to find the
minimum value of the function $\Veff(\sigma_i,\varphi_i)$, it is
sufficient to restrict ourselves to the configuration of the
variables with, e.g., $\varphi_1=\varphi_2=0$, $\rho=\sigma_1$. The
stationarity (gap) equations for the effective potential
(\ref{3.12}) with respect to the independent variables
$M_{1,2}=|\sigma_2\pm\sigma_1|$ then read
\begin{eqnarray}
\frac{\partial \Veff}{\partial M_i}=M_i\left
(\frac{g}{4v_{\rm F}}+\frac{M_i}{2\pi v^2_{\rm F}}\right )=0,~~~~i=1,2~.
\label{3.13}
\end{eqnarray}
It is evident from eq. (\ref{3.13}) that for subcritical values of
the bare coupling constant $G<G_c$ or, equivalently, at $g>0$ the
gap equations (\ref{3.13}) have only a trivial solution,
$M_{1,2}=0$. It corresponds to the symmetrical global minimum point
of the effective potential (\ref{3.12}), $\langle\sigma_i\rangle
=\langle\varphi_i\rangle =0$. However, for the supercritical values
of the bare coupling constant $G>G_c$ or, equivalently, at $g<0$,
the obvious solution of the gap equations (\ref{3.13}) is
$M_1=M_2=-\pi g v_{\rm F}/2$. Thus, expressed in terms of the
variables $\sigma_{1,2}$ at $\varphi_{1,2}=0$, in this case there
are two different global minimum points of the effective potential
(\ref{3.12}) corresponding to (i) $\langle\sigma_1\rangle =-\pi g
v_{\rm F}/2$, $\langle\sigma_2\rangle =\langle\varphi_1\rangle
=\langle\varphi_2\rangle =0$ and (ii) $\langle\sigma_2\rangle =-\pi
g v_{\rm F}/2$, $\langle\sigma_1\rangle =\langle\varphi_1\rangle
=\langle\varphi_2\rangle =0$, i.e. the minimum value of the
effective potential is degenerated. Note also that in a similar way
the smallest value of the function (\ref{3.12}) can be investigated
equivalently in terms of other two variable configurations,
$\sigma_1=\varphi_2=0$, $\rho=\varphi_1$ and $\varphi_1=\sigma_1=0$,
$\rho=\varphi_2$. As a result, one can easily find that in addition
to global minimum points of $\Veff(\sigma_i,\varphi_i)$
corresponding to the two above mentioned condensate structures (i)
and (ii), there are two another global minimum points corresponding
to (iii) $\langle\varphi_1\rangle =-\pi g v_{\rm F}/2$,
$\langle\sigma_1\rangle =\langle\sigma_2\rangle
=\langle\varphi_2\rangle =0$ and (iv) $\langle\varphi_2\rangle =-\pi
g v_{\rm F}/2$, $\langle\sigma_1\rangle =\langle\sigma_2\rangle
=\langle\varphi_1\rangle =0$. It is evident that all global minimum
points, corresponding to condensate structures (i),$\dots$,(iv), are
degenerated. However, they correspond to ground states of different
phases of the model. Their symmetry properties are described in
Table \ref{Tab1} and just after it.

\subsubsection{Exciton spectrum}

It is further instructive to exhibit also the mass spectrum of
bound-state excitons. Suppose that the ground state of the system
described by the Lagrangians (\ref{2.35}) and (\ref{2.38}) is
determined by condensate values $\langle\sigma_k\rangle$,
$\langle\varphi_k\rangle$ ($k=1,2$). To find the masses of excitons
in this ground state, one should perform in the effective action
$S_{\rm eff}(\sigma_i,\varphi_i)$ of eq. (\ref{3.2}) a shift around
condensates (mean field values),
$\sigma_k(x)\to\langle\sigma_k\rangle+\sigma_k(x)$,
$\varphi_k(x)\to\langle\varphi_k\rangle+\varphi_k(x)$ ($k=1,2$),
where the quantities $\sigma_k(x)$, $\varphi_k(x)$ are now
fluctuating fields. Then it is necessary to take into account the
fact that the obtained effective action
$S_{\rm eff}(\sigma_i,\varphi_i)$ is a generating functional of
one-particle irreducible (1PI) Green functions of the fluctuating
fields.

To be more specific, let us here consider the phase  with
$\langle\sigma_1\rangle\equiv m_1v_{\rm F}^2\sim \langle\bar\psi\psi\rangle\ne
0$, $\langle\sigma_2\rangle =\langle\varphi_1\rangle
=\langle\varphi_2\rangle =0$.
For the particular case $G_k=H_k=G$
($k=1,2$), considered in the previous section 3.1.1, this phase
corresponds to a global minimum point (i) of the effective potential
(\ref{3.12}).
However, this phase is also realized for other, not so
trivial as in the section 3.1.1, relations between coupling
constants of the model (\ref{2.35}), (\ref{2.38})  \cite{n13,n27}.
(See also Appendix B for the structure of the condensates in the
most general case. In particular, it follows from Table \ref{T1} that $\langle\sigma_1\rangle=-\pi g_1v_{\rm F}/2$.) Performing in this case a simplest field shift in the effective action (\ref{3.2}), $\sigma_1(x)\to\langle\sigma_1\rangle+\sigma_1(x)$, we obtain the two point 1PI Green functions (inverse propagators) of the
fluctuating fields,
\begin{eqnarray}
\Gamma_{\phi_k\phi_k}(x-y)=\frac{\delta^2S_{\rm eff}}{\delta\phi_k(x)\delta\phi_k(y)}\Big
|_{\sigma_i,\varphi_i=0}~~,~\phi_k=\{\sigma_1,\sigma_2,\varphi_1,\varphi_2\}.
\label{3.16}
\end{eqnarray}
It then follows from eqs. (\ref{3.2}) and (\ref{3.16}) that
\begin{eqnarray}
\Gamma_{\phi_k\phi_k}(x-y)=-\frac{1}{2v_{\rm F}G_{\phi_k}}\delta^{(3)}(x-y)+i{\rm
Tr}_{s}\left [\hat t_kG_0(x-y)\hat t_k G_0(y-x)\right ], \label{3.17}
\end{eqnarray}
where we use the notations
\begin{eqnarray*}
G_{\phi_k}=\left\{G_1,G_2,H_1,H_2\right\}~,~~~~~\hat
t_k=\left\{{\rm
I}_4,\gamma^{35},i\gamma^5,\gamma^3\right\}~,~~~~k=1,...,4,
\end{eqnarray*}
and $G_0(x-y)$ is the inverse of the operator
$(i\lefteqn{/}\widetilde{\partial}-\langle\sigma_1\rangle)$
with no flavor indices (recall that its mass term is
$\langle\sigma_1\rangle=-\pi g_1v_{\rm F}/2\equiv m_1v_{\rm F}^2$).
It has in the spinor space the following matrix elements
\begin{eqnarray}
G_0(x-y)_{\alpha\beta}=\int\frac{d^3p}{(2\pi)^3}\left
(\frac{1}{\lefteqn{/}\widetilde p-m_1v_{\rm F}^2}\right
)_{\alpha\beta}\e^{-ip(x-y)}, \label{3.18}
\end{eqnarray}
where $\widetilde p=(p^0,v_{\rm F}\vec p)$, and for spinor indices
we have $\alpha,\beta =1,...,4$.
The straightforward loop calculations, required by eq(3.16),
can be found in Appendix C. In momentum space (Minkowski metric),
we then obtain
\begin{eqnarray}
\Gamma_{\sigma_1\sigma_1}(p)=
\frac{\widetilde p^2 - (2m_1v_{\rm F}^2)^2}
{2\pi v_{\rm F}^2 \sqrt{-\widetilde p^2}}\Gamma(p),&~~&\Gamma(p)=\tan^{-1}\left (\frac{\sqrt{-\widetilde p^2}}{2m_1v_{\rm F}^2}\right ),\nonumber\\
\Gamma_{\sigma_2\sigma_2}(p)=-\frac{1}{2v_{\rm F}}(g_2-g_1)+\frac
{\widetilde p^2-(2m_1v_{\rm F}^2)^2}{2\pi v_{\rm F}^2 \sqrt{-\widetilde
p^2}}\Gamma(p),&~~
&\Gamma_{\varphi_k\varphi_k}(p)=-\frac{1}{2v_{\rm F}}(h_k-g_1)-\frac{\sqrt{-\widetilde
p^2}}{2\pi v_{\rm F}^2 }\Gamma(p). \label{3.19}
\end{eqnarray}
To obtain these expressions, the gap equations (\ref{3.11}) and the
relations (\ref{3.10}) between bare couplings $G_i$, $H_i$ and
finite parameters $g_i$, $h_i$ have been used.
It is worth emphasizing that in the case $v_{\rm F}=c=1$ our
graphen-like expressions in eq. (\ref{3.19}) coincide with the QED results
obtained in refs. [15,17].

The inverse expressions of eq. (\ref{3.19}) are just the exciton
propagators, the singularities of which determine their mass
spectrum and dispersion laws, i.e. the relations between their energies and spatial momenta. Thus, the scalar excitation $\sigma_1$ corresponds to a stable particle with a mass $m_{\sigma}=2m_1$. The quasiparticle
$\sigma_2$ is a scalar resonance, corresponding to a pole of the
propagator on the second sheet of its region of analyticity. The
fields $\varphi_1$, $\varphi_2$ correspond to two scalar and
pseudoscalar stable bound states of two fermions
(see eq. (\ref{2.39}) and Table 1) with nonzero binding energy. To
obtain the 1PI two-point Green functions in the phase, where only
$\langle\sigma_2\rangle\sim \langle\bar\psi\gamma^{35}\psi\rangle\ne
0$, it is required to make the replacements
$\sigma_1\leftrightarrow\sigma_2$, $g_1\leftrightarrow g_2$ in eq.
(\ref{3.19}) etc.

It is evident that under certain restrictions of coupling constants,
the considered model Lagrangian (\ref{2.35}) acquires additional
continuous symmetries. For illustrations, let us consider the
effective potential (\ref{3.9}) and assume there the condition
$\{g_1=h_1=g<0,~g_2, h_2>g\}$, where the global minimum point of
$\Veff$ corresponds to $\langle\sigma_1\rangle\sim
\langle\bar\psi\psi\rangle\ne 0$, $\langle\sigma_2\rangle
=\langle\varphi_1\rangle =\langle\varphi_2\rangle =0$ (see Appendix
B). In this case the Lagrangian (\ref{2.35}) is invariant under
continuous chiral transformations
\begin{eqnarray}
U_{\gamma^5}(1)~:~~\psi\to\exp (i\alpha\gamma^5)\psi, \label{3.20}
\end{eqnarray}
where
\begin{eqnarray}
(\bar\psi\psi)\to (\bar\psi\psi)\cos 2\alpha+(\bar\psi
i\gamma^5\psi)\sin 2\alpha,&~~&(\bar\psi i\gamma^5\psi)\to
-(\bar\psi\psi)\sin 2\alpha+(\bar\psi i\gamma^5\psi)\cos 2\alpha,\nonumber\\
(\bar\psi\gamma^3\psi)\to
(\bar\psi\gamma^3\psi),&~~&(\bar\psi\gamma^{35}\psi)\to
(\bar\psi\gamma^{35}\psi). \label{3.21}
\end{eqnarray}
It then follows from eq. (\ref{3.19}) that the propagator of the
pseudoscalar field $\varphi_1$ has a pole at $\tilde p^2=0$, i.e. in
the mean field (large $N_{\rm f}$) approximation spontaneous breakdown
(SB) of chiral symmetry with an associated massless Goldstone boson
(GB) occurs. An analogous situation happens in the case $g_1=h_2$
for a continuous transformation with $\gamma^3$-generator. Note that
the above discussion of SB of a continuous symmetry with associated
appearance of a GB was based on the mean field approximation
neglecting finite $1/N_{\rm f}$ corrections. On the other hand, for finite
temperature, there exists the important Mermin-Wagner-Coleman (MWC)
no-go theorem \cite{30} which for (2+1)-dimensional systems forbids
SB of a continuous symmetry. In principle, this requires to go
beyond the dominant expressions in the $1/N_{\rm f}$ expansion and to take
into account phase fluctuations of bound-state fields, related to
vortex excitations and the Kosterlitz-Thouless transition
\cite{28,31}. The consideration of related $1/N_{\rm f}$ corrections is,
however, outside the scope of this paper. By this reason, our
numerical investigations of phase transitions at finite $T$ and
$\mu$ in Section 4 will be restricted to a simpler GN-type
interaction with discrete chiral symmetry $\gamma^5$, where the
MWC-theorem does not hold.

\subsection{Nanotubes from hexagonal lattice}
\subsubsection{Boundary conditions}

For possible applications to nanotubes in magnetic fields, we shall
now investigate the case where one spatial direction is
compactified
and the (hexagonal lattice) sheet is rolled up to a cylinder. In
particular, we shall consider the cylinder as a sort of a (2+1)D
brane embedded in flat (3+1)D space-time. Fermions living in the
brane are then moving under the influence of a parallel homogeneous
magnetic field directed along the cylinder axis. Here and in what
follows we use, in the bulk, either Cartesian or cylindrical
coordinates $(x,y,z)$ or $(\rho, \varphi, z)$,  respectively, and
coordinates $(x^1,x^2)$ on the cylinder surface with $x^1$ pointing
in the z-direction, and $x^2=R\varphi$ being the compactified
coordinate, which has a length $L=2\pi R$ with $R$ as the cylinder
radius. The basis vectors on the cylinder surface are ${\bf
e}_1\equiv {\bf e}_z$  and ${\bf e}_2\equiv {\bf e}_\varphi$. The
$z$-axis in the bulk is parallel to the cylinder axis, and the
vector potential associated to a magnetic field
$B_0$
parallel to the
cylinder ($z$)-axis is given by
$\vec A=\frac{\rho}{2}B_0 {\bf
e}_\varphi$
in the bulk and as
$\vec {\cal
  A}=\frac{R}{2}B_0 {\bf e}_\varphi$
  on the cylinder surface. This field
$\vec {\cal A}$ has to be included in the form of a covariant
derivative by replacing $\partial_2\rightarrow
D_2=\partial_2+ie{\cal
  A}_2$ in the Lagrangian. Alternatively, one might also keep
$\partial_2$ and include an effective magnetic phase $\phi$,
\begin{equation}
\phi=\frac{e{\cal A}_2 L}{2\pi}=\frac{\Phi_m}{\Phi_m^0}
\label{3.22}\end{equation} into the boundary condition of the
fermion field $\psi(x^0,x^1,x^2)$
\begin{equation}
\psi(x^0,x^1,x^2+L)=\e^{2\pi i (\phi+\alpha)}\psi(x^0,x^1,x^2)
.\label{3.23}\end{equation} Here $\Phi_m$ is the magnetic flux
passing through the tube cross section, $\Phi_m^0=2\pi/e$ is the
magnetic flux quantum and $\alpha$ is determined by the lattice
structure (cf. eq. (\ref{3.24})). Notice that the above type of
interaction with a magnetic flux $\Phi$ inserted via ${\cal A}_2$ in
the covariant derivative $D_2$ can just be considered as a
manifestation of the Aharonov-Bohm (AB) effect \cite{n33}. At the
same time, an external bulk magnetic field $B_{\parallel}$ included
in the direction of the cylinder axis and existing at the cylinder
surface should also lead to a Zeeman spin-interaction. In fact, the
magnetic moment from the real spin of fermions will interact with
the surface magnetic field $B_\parallel$. Such a Zeeman term has
then to be added as in eq. (\ref{2.37}).

Taking into account the properties of the hexagonal graphene-like
lattice in nanotubes, it has been shown that the fermion field
$\psi(t,\vec r)$ satisfies at the $K$-point the following boundary
condition \cite{n33,n34}
\begin{equation}
\psi_K(x^0,\vec r+\vec L)=\e^{2\pi
i(\phi-\frac{1}{3}\nu)}\psi_K(x^0,\vec r)
,\label{3.24}\end{equation} where $\nu=(0,\pm 1)$, and the phase
$\phi$ is given in eq. (\ref{3.22}). An analogous expression follows
for $\psi_{K'}$ with the replacement $\nu\rightarrow-\nu$,
\begin{equation}
\psi_{K'}(x^0,\vec r+\vec L)=\e^{2\pi
i(\phi+\frac{1}{3}\nu)}\psi_{K'}(x^0,\vec r)
.\label{3.25}\end{equation}

The field spinors satisfying the boundary conditions (\ref{3.24}),
(\ref{3.25}) can be written as Fourier decomposition
\begin{equation}
\psi=
\begin{pmatrix}
\psi_K^A\\
\psi_K^B\\
-i\psi_{K'}^B\\
i\psi_{K'}^A
\end{pmatrix}
=\frac{1}{L}\sum\limits_{n=-\infty}^{\infty}\e^{i\left[\frac{x^2}{R}(n+\phi)+p_1x^1+p_0x^0\right]}
\begin{pmatrix}
\psi_{Kn}^{(1)}\\
\psi_{K'n}^{(2)}\\
\end{pmatrix}
,\label{3.26}\end{equation} where
\begin{equation}
\begin{gathered}
\psi_{Kn}^{(1)}=
\begin{pmatrix}
\psi_{Kn}^{A}\\
\psi_{Kn}^{B}
\end{pmatrix}
\e^{-i\frac{x^2}{R}\left(\frac{\nu}{3}\right)} \\
\psi_{K'n}^{(2)}=
\begin{pmatrix}
-i\psi_{K'n}^{B}\\
i\psi_{K'n}^{A}
\end{pmatrix}
\e^{i\frac{x^2}{R}\left(\frac{\nu}{3}\right)}.\label{3.27}
\end{gathered}
\end{equation}
This then leads for $\nu\neq 0$ to a nonvanishing ("semiconductor")
gap $\Delta\cal E$ between the conduction and valence bands of
non-interacting fermions with vanishing dynamical masses. Indeed,
the azimuthal components of the $p_2$ momentum are
\begin{equation}
p_{\nu\phi}(n)=\frac{2\pi}{L}(n+\phi-\frac{\nu}{3})
,\label{3.28}\end{equation}
so that
\begin{equation}
\Delta{\cal E}(n=\phi=p_1=0)=v_{\rm
F}\frac{4\pi}{L}\frac{|\nu|}{3}\neq 0 .\label{3.29}\end{equation}

On the other hand, $\Delta {\cal E}=0$ if $\nu=0$, and one gets
"metallic" behavior \cite{n33,n34}. Obviously, such an energy gap
increases in the insulator phase with a dynamical mass gap $m$,
where one has
\begin{equation}
{\cal E}^\pm(p_1,p_{\nu\phi}(n)) =\pm\sqrt{v_{\rm F}^2p_1^2+v_{\rm
    F}^2p_{\nu\phi}^2(n)+
(mv_{\rm F}^2)^2},\label{3.30}\end{equation} and thus
\begin{equation}
\Delta{{\cal E}} (n=p_1=\phi=0)=2\sqrt{v_{\rm
F}^2\left(\frac{2\pi}{L}\right)^2\left(\frac{\nu}{3}\right)^2+(mv_{\rm F}^2)^2}
.\label{3.31}\end{equation}

\subsubsection{Thermodynamic potential for nanotubes}

Let us finally consider the thermodynamic potential $\Omega_T$ for a
nanotube at finite temperature and particle density. In this case,
we have to calculate the $\rm Tr$-operation in eq. (\ref{3.2}) by
taking into account the Fourier decomposition (\ref{3.26}) of
spinors. In order to get the corresponding (unrenormalized)
thermodynamic potential $\Omega_T$, one has to perform a replacement
of the $p_0$-integration in eq. (\ref{3.5}) by a summation over
Matsubara frequencies $\omega_\ell$, following the rule
\begin{equation}\begin{gathered}
\int\limits_{-\infty}^{\infty}{\frac{dp_0}{2\pi}}f(p_0)\to\frac{i}{\beta}\sum\limits_{\ell=-\infty}^{\infty}f(i\omega_\ell),
\label{3.32} \end{gathered}\end{equation} where
$\omega_\ell=\frac{2\pi}{\beta}\left(\ell+\frac{1}{2}\right),~~~\ell=0,\pm
1, \pm 2,...$ and $\beta=\frac{1}{T}$ is the inverse temperature.
Next, the chemical potential $\mu$ will be introduced by the
standard shift $\omega_\ell\to \omega_\ell-i\mu$
\footnote{After taking temperature and chemical potential into
account, the effective potential of the model acts as a
thermodynamic potential. However, in order to keep consistency in
different sections of the present paper, we shall continue to denote
it as $\Veff$.}.
Finally, we have to take into account the boundary condition of the nanotube which requires to replace the momentum component $p_2$ by the
expression (\ref{3.28}), $p_2\to p_{\nu\phi}(n)
=\frac{2\pi}{L}(n+\phi-\frac{\nu}{3})$, where the phase $\phi$ is
expressed by the magnetic AB-flux (see eq. (\ref{3.22})). A lengthy
but straightforward calculation then gives
\begin{equation} \begin{gathered}
\Veff(\sigma_i,\varphi_i,T,\hat \mu,\phi)=
\sum\limits_{k=1}^{2}\left\{
\left(\frac{\sigma_k^2}{4v_{\rm F}G_k}+\frac{\varphi_k^2}{4v_{\rm
F}H_k}\right) -\frac{1}{\beta L}\sum\limits_{s=\pm
1}\sum\limits_{\ell=-\infty}^{\infty}\sum\limits_{n=-\infty}^{\infty}
\int{\frac{dp_1}{2\pi}} \ln\left[
\left(\frac{2\pi}{\beta}\left(\ell+\frac{1}{2}\right)-i\hat
\mu\right)^2+ \right.\right.
\\
\left.\left. +v_{\rm F}^2\left(\frac{2\pi}{L}\right)^2\left(n+\phi
-\frac{\nu}{3}\right)^2+ v_{\rm F}^2 p_1^2 + M_k^2 \right] \right\}
,\label{3.33}\end{gathered}\end{equation} where we have included the
Zeeman term into the effective chemical potential $\mu$ according to
eq. (\ref{2.37}),
\begin{equation}
\hat \mu = \mu-\frac{g}{2}s\mu_{\rm B}B_\parallel
\label{3.34}\end{equation}
with $s=\pm 1$ for up/down spin, and the mass gaps $M_k$ are given after eq.(\ref{3.5}). Eq. (\ref{3.33}) is one of the main results of this paper.

In the next Section we shall use $\Veff$ for a numerical
investigation of the chiral phase transition in the $(\beta,L)$ and
$(\mu, T)$ plane in dependance on the magnetic flux $\phi$. For a
first application and as illustration, we shall restrict us now to a
simple model version of eq. (\ref{3.33}) by choosing coupling constants
in such a way that the mean field solutions  are
$\sigma_1=\sigma\neq 0$, $\sigma_2=\varphi_1=\varphi_2=0$.
Neglecting temporarily the Zeeman effect, putting $\hat \mu \to
\mu$, we then have ($G_1\to G$)
\begin{equation} \begin{gathered}
\Veff(\sigma,T,\mu,\phi)= \frac{\sigma^2}{4v_{\rm F}G}
-\frac{2}{\beta
L}\sum\limits_{\ell=-\infty}^{\infty}\sum\limits_{n=-\infty}^{\infty}
\int{\frac{dp_1}{2\pi}} \ln\left[
\left(\frac{2\pi}{\beta}\left(\ell+\frac{1}{2}\right)-i
\mu\right)^2+
\right. \\
\left.\left. +v_{\rm F}^2\left(\frac{2\pi}{L}\right)^2\left
(n+\phi-\frac{\nu}{3}\right)^2+ v_{\rm F}^2 p_1^2 + \sigma^2 \right]
\right\} .\label{3.35}\end{gathered}\end{equation} Note that
expression (\ref{3.35}) generalizes the results of the compactified
Gross--Neveu (GN) models \cite{n28,n29,n32,n34} to finite $T$ and $
\mu$. Recall also that when investigating the thermodynamic potential
(\ref{3.35}) of the compactified GN-model, the MWC-theorem \cite{30}
does not apply, since one considers the spontaneous breakdown and
restoration of a discrete chiral $\gamma^5$-symmetry at finite $T$.

\section{Chiral phase transitions in nanotubes \label{sec4}}
\subsection{Phase structure in nanotubes with non-zero chemical potential}

As an illustration of the above general results, we will now
investigate the phase transitions in the model described by eq.
(\ref{3.35}), omitting, for simplicity, the influence of the Zeeman
effect and taking further $\nu=0$ (metallic case), but considering
finite temperature and chemical potential. We will, however,  keep
the Aharonov-Bohm phase $\phi$ untouched for further calculations.
After subtraction of terms independent of $\sigma$ that do not
affect the symmetry properties, we write
\begin{equation}
\Veff= \frac{\sigma^{2}}{4v_{\rm F}G}-\frac{2} {\beta L}
\sum_{\ell=-\infty}^{\infty}\sum_{n=-\infty}^{\infty}\int\frac{dp_{1}}{2\pi}
\ln\left[1+\frac{\sigma^{2}}{(\frac{2\pi}{\beta} (\ell+\frac{1}{2})-
i\mu)^{2}+(v_{\rm F}\frac{2\pi}{L})^{2}( n+\phi)^{2}+
v_{\rm F}^2p_{1}^{2}}\right] .\end{equation} Further we use the following
formula:
\begin{equation}\begin{split}
\sum_{\ell=-\infty}^{\infty}\ln(1+\frac{b^{2}}{(\ell+\alpha)^{2}+a^{2}})=
\sum_{l=-\infty}^{\infty}(\ln((\ell+\alpha)^{2}+a^{2}+b^{2})-\ln((\ell+\alpha)^{2}+a^{2}))
\label{prim1}
\\
=\int_{-\infty}^{\infty} d\tau \ln(1+\frac{b^{2}}{\tau^{2}+a^{2}})+
\ln\frac{1-2\cos(2\pi\alpha)
e^{-2\pi\sqrt{a^{2}+b^{2}}}+e^{-4\pi\sqrt{a^{2}+b^{2}}}}{1-2\cos(2\pi\alpha)
e^{-2\pi\sqrt{a^{2}}}+e^{-4\pi\sqrt{a^{2}}}}
,\end{split}\end{equation} which can be found e.g. in eq. \cite{n28},
and obtain
\begin{equation}\begin{gathered}
\Veff=\frac{\sigma^{2}}{4v_{\rm F}G}-\frac{2} {
L}\sum_{n=-\infty}^{\infty}\int\frac{dp_{1}}{2\pi}\frac{dp_{0}}{2\pi}
\ln\left[1+\frac{\sigma^{2}}{(v_{\rm F}\frac{2\pi}{L})^{2}
(n+\phi)^{2}+v_{\rm F}^2p_{1}^{2}+p_{0}^{2}}\right]-
\\
-\frac{2} {\beta L
}\sum_{n=-\infty}^{\infty}\int\frac{dp_{1}}{2\pi}\ln\left[1+2\text{ch}(\beta
\mu)\exp\left(-\beta E_{n,p_1}\right) +\exp\left(-2\beta
E_{n,p_1}\right)\right],
\end{gathered}\end{equation}
where $E_{n,p_1}=\sqrt{
v_{\rm F}^2p_1^2+\left(v_{\rm F}\frac{2\pi}{L}\right)^{2}(n+\phi)+\sigma^{2}}$.
To transform the second summand in $\Veff$, we use eq.
(\ref{prim1}). The result is $\Veff=V_{(0)}+\VL+\VmT$, where
\begin{equation}\label{www1}
\Vo\equiv\frac{\sigma^{2}}{4v_{\rm F}G}-2\int\frac{ d^{3}p}
{(2\pi)^{3}}
\ln\left(p_0^{2}+v_{\rm F}^2 p_1^2+v_{\rm F}^2 p_2^2+\sigma^{2}\right)=
\frac{1}{\pi v_{\rm
F}^2}\left(\frac{|\sigma|^{3}}{3}-\frac{\sigma^{2}\sigma_{0}}{2}\right)
,\;\;\; \sigma_0=-{\pi v_{\rm F} g/2},
\end{equation}

\begin{equation}\label{www2}\begin{gathered}
\VL= -\frac{2}{L}\int \frac{d^{2}p}{\left(2\pi\right)^{2}} \ln
\frac{1-2\cos\left(2\pi \phi\right)\e^{-\frac{L}{v_{\rm
F}}\sqrt{p_0^{2}+v_{\rm F}^2 p_1^2+\sigma^{2}}}+\e^{-2\frac{L}{v_{\rm
F}}\sqrt{p_0^{2}+v_{\rm F}^2 p_1^2+\sigma^{2}}}}{1-2\cos\left(2\pi
\phi\right)\e^{-\frac{L}{v_{\rm F}}\sqrt{p_0^{2}+v_{\rm F}^2
p_1^2}}+\e^{-2\frac{L}{v_{\rm F}}\sqrt{p_0^{2}+v_{\rm F}^2 p_1^2}}}=
\\
=\frac{2 v_{\rm F}}{\pi L^{3}}\sum_{n=1}^{\infty}\frac{\cos(2\pi
\phi n)}{n^{3}}\e^{-\frac{L\sigma n}{v_{\rm F}}}+ \frac{2\sigma}{\pi
L^{2}}\sum_{n=1}^{\infty}\frac{\cos(2\pi \phi
n)}{n^{2}}\e^{-\frac{L\sigma n}{v_{\rm F}}}=
\\
=2\Re \e\left(\frac{v_{\rm F}}{\pi L^{3}}\Li_{3}(\e^{-\frac{L}{v_{\rm
F}}\sigma+2\pi i \phi}) +\frac{\sigma}{\pi L^{2}}\Li_{2}(e^{
-\frac{L}{v_{\rm F}}\sigma+2\pi i\phi}) \right) ,
\end{gathered}\end{equation}
\begin{equation}\label{www3}\begin{gathered}
\VmT= -\frac{2} {\beta
L}\sum_{n=-\infty}^{\infty}\int\frac{dp_{1}}{2\pi}\ln\left(1+2\cosh\left(\beta
\mu\right)\exp\left(-\beta E_{n,p_1}\right) \exp\left(-2\beta
E_{n,p_1}\right)\right)
\\
=-\frac{2} {\beta L}\sum_{n=-\infty}^{\infty}\int\frac{dp_{1}}{2\pi}
\left[ \ln\left(1+e^{-\beta\left(E_{n,p_1}^{+}\right)}\right)+
\ln\left(1+\e^{-\beta\left(E_{n,p_1}^{-}\right)}\right) \right]
,\end{gathered}\end{equation} and
$E_{n,p_1}^{\pm}=E_{n,p_1}\pm\mu$.
Note that the quantity $\Vo$ in eq. (\ref{www1}) is the effective
potential of the system in vacuum, i.e. at $T=0$ and $\mu=0$. Hence,
its final expression in the right hand side of (\ref{www1}) can be
obtained in the same way as in Sect. 3.1 (compare eq. (\ref{www1})
with the more general expression (\ref{3.9})). So, the
renormalization invariant coupling constant $g$ in eq. (\ref{www1})
is a particular case of relations (\ref{3.10}), $g=1/G-1/G_c$. At
$T=0$, $\mu=0$ and $g<0$ the parameter $\sigma_0$ in eq.
(\ref{www1}) is the nonzero vacuum expectation value of the $\sigma$
field, $\sigma_0=\langle\sigma (x)\rangle$, which corresponds to
spontaneous breaking of $\gamma^5$ chiral symmetry. However, at
$g>0$ we have in this case $\langle\sigma (x)\rangle=0$ and intact
$\gamma^5$ symmetry. Moreover, $\Li_{s}$ are polylogarithms (here
di- and trilogarithms), while $\Re e$ means the real part of the
expression. Further numerical calculations resulted in the graphs
shown below. For instance, Fig. \ref{Fig.muT} demonstrates the phase
structure of the model in the plane $(T,\mu)$ with
$\phi=0.05\approx0$ (an actual numerical computation for $\phi=0$
gives nearly the same result, but requires much more time for
drawing accurate plots). The temperature is given in units of the
critical temperature $T_c=\frac{1}{\beta_c}=\frac{\pi |g| v_{\rm
F}}{4\ln(2)}$ and the chemical potential is given in units of
$\mu_c=\pi |g| v_{\rm F}/2$, where $g=g_1$ is defined in eq.
(\ref{3.101}). These values correspond to the temperature and
chemical potential that restore symmetry in a "flat" model without
any other external parameters. The circumference of the compactified
dimension is chosen small enough to make the model differ from the
"flat" one ($L=1.2L_c$, where $L_c=v_{\rm F}\beta_c$). The resulting
structure is similar to what has been found in ref. \cite{n2c} for a 2D
GN model. The result can be interpreted as a manifestation of
dimensional reduction in the asymmetrical phase, which exists for
low enough temperature and periodic boundary conditions (see ref.
\cite{202}).  Areas I and III in Fig. 3 correspond to broken
symmetry. The difference  between them is that in area I only one
minimum of $ \Veff$ exists and  is located at $\sigma\neq0$, whereas
in area III there are two minima. Additionally to the one presented
in area I, there is a local minimum at  $\sigma=0$, and a global one
at $\sigma\neq0$.  Areas II and IV are areas of restored symmetry.
In area II only the  trivial minimum at $\sigma=0$ exists, while in
area IV there are two local minima of the effective potential, the
trivial one and a non-trivial one at $\sigma\neq 0$, however the
global minimum is trivial. Hence the line $BE$ in Fig. 3 is the line
of a phase transition of the first kind, whereas the line $AB$ is
the line of a phase transition of the second kind. Lines $BD$ and
$BC$ are not  lines of  phase transitions, but are lines, where the
trivial and non-trivial minima vanish/appear as  local minima.

\begin{figure}\label{gr}
\begin{center}
\includegraphics[height=200pt]{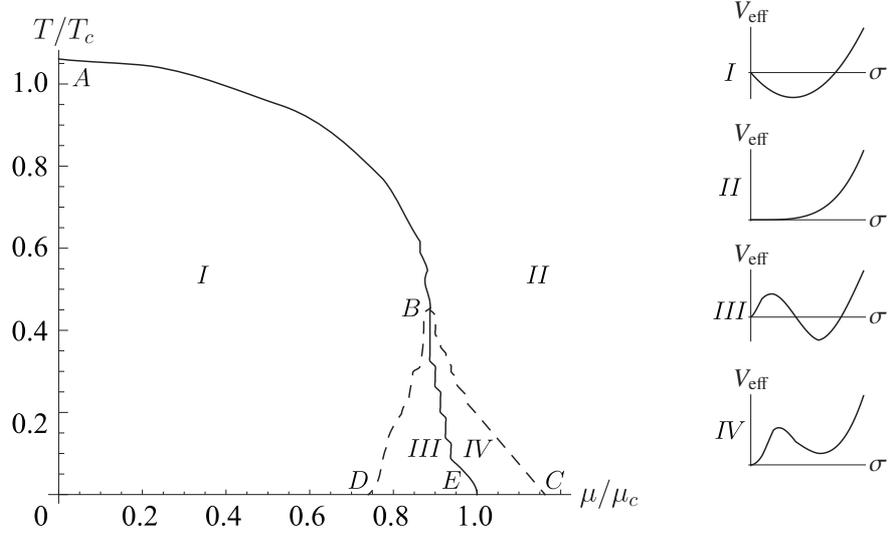}
\caption{\it{Phase diagram of the model under the influence of
    chemical potential and temperature. Figures in the right
    schematically show the behavior of the effective potential
    $\Veff$ as a function of $\sigma$. The explanation of the phases I-IV is given in the text below.}}\label{Fig.muT}
\end{center}
\end{figure}

Obviously, when the Zeeman effect is taken into account by the
replacement $\mu\to\hat\mu$, only the summand $\VmT$ needs to be
modified, leading to
\begin{equation}\label{wwwZ}\begin{gathered}
\VmT^Z=-\frac{1} {\beta
L}\sum_{n=-\infty}^{\infty}\int\frac{dp_{1}}{2\pi} \left[
\ln\left(1+\e^{-\beta\left(E_{n,p_1}\right)^{+}_{\ua}}\right)+
\ln\left(1+\e^{-\beta\left(E_{n,p_1}\right)^{-}_{\ua}}\right)+
\right.
\\
\left. +
\ln\left(1+\e^{-\beta\left(E_{n,p_1}\right)^{+}_{\da}}\right)+
\ln\left(1+\e^{-\beta\left(E_{n,p_1}\right)^{-}_{\da}}\right) \right]
,\end{gathered}\end{equation} where
$(E_{n,p_1})^{\pm}_{\da\ua}=E_{n,p_1}\pm\mu_{\da\ua}$ and
$\mu_{\da\ua}$ represents $\hat \mu$ for $s=\pm 1$. Symmetry between
$\mu$ and $\delta\mu=\frac{g}{2}\mu_{\rm B}B_\parallel$ can now be
noticed. In fact, the replacement $\mu\leftrightarrow\delta\mu$ does
not change the effective potential. Hence Fig.~\ref{Fig.muT} can
also be interpreted as the phase structure of the system under the
influence of the Zeeman effect but with a vanishing chemical
potential.

\subsection{Phase structure in nanotubes under the influence of
  the Aharonov--Bohm effect}

Here we will investigate the influence of the Aharonov--Bohm effect
on the symmetry properties of the nanotubes
at zero chemical potential.
>From the discussion after eq. (\ref{3.22}) it is, in particular,
clear that the Aharonov--Bohm effect represents the change in the
periodicity condition for the nanotubes.

Let us put $\mu=0$ in the general formula (\ref{3.35}) and calculate
$\Veff(\sigma, \phi, T, \mu)|_{\mu=0}$.
By using the proper time representation:
\begin{equation}\ln \frac{A}{B}=-\int\limits_0^\infty \cfrac{ds}{s}\phantom{a}\left(\exp(-sA)-\exp(-sB)\right)\end{equation}
and Poisson resummation formula
\begin{equation}\label{Poison_resum}
\sum\limits_{\ell=-\infty}^{+\infty}\exp\left[-s\left(\cfrac{2\pi
\ell}{B}+C\right)^2\right]= \cfrac{B}{2\sqrt{\pi s}}\left[ 1+
2\sum\limits_{\ell=1}^{+\infty}\exp\left(-\cfrac{B^2
\ell^2}{4s}\right)\cos(BC\ell) \right],\end{equation} we write the
effective potential in the form
\begin{equation}\begin{gathered}
\Veff(\sigma, \phi, T)\,=\, \cfrac{\sigma^2}{4v_{\rm F}G}
+\cfrac{1}{4\pi^{3/2} v_{\rm F}^2}\int\limits_0^\infty
\cfrac{ds}{s^{5/2}} \phantom{a} \left[\exp(-s\sigma^2)\right]\times
\\
\times \left[1+
2\sum\limits_{\ell=1}^{+\infty}(-1)^\ell\exp\left(-\cfrac{\beta^2
\ell^2}{4s}\right)\right] \times \left[ 1+
2\sum\limits_{n=1}^{+\infty}\exp\left(-\cfrac{L^2 n^2}{4sv_{\rm
F}^2}\right)\cos(2\pi n \phi)\right]+{\rm c.t.},
\label{4.10}\end{gathered}\end{equation}
where the counterterm $\rm c.t.$ does not depend on the parameters to be investigated and is further omitted.
Various combinations of products of summands in the last two factors
of the above formula correspond to  different physical situations.
By considering only summands that are equal to unity, we can
investigate a ``flat''\,(2+1)D GN model without compactification,
temperature and external fields
($L\to\infty,\,T=0,\,\mathcal{A}_2=0$). The term containing
$\beta=1/T$ in the exponent of the first cofactor gives the
temperature dependence, while the summand containing $L$ in the
exponent of the second cofactor is due to compactification of
spatial dimension (with respect to boundary conditions). One of the
key characteristics of the GN model is the restoration of chiral
symmetry ($\sigma \to 0$) under the influence of high temperature.
In this section we aim at the study of the role of all the
parameters ($L,\,T,\,\mathcal{A}_2$), while in ref. \cite{n29} this model
was investigated at $T=0$.
The consideration of all the summands along with their product
allows to consider now the case of finite temperature and  spatial
compactification at the same time. The entire effective potential is
the sum of all its parts
\begin{equation}\label{VeffAsASum}
\Veff=\Vo+V_{(T)}+\VL+V_{(\times)},
\end{equation}
where $V_{(0)}, V_{(T)}$, and $V_{(L)}$
correspond to zero
temperature, finite temperature and compactification parts,
respectively, while the cross-term $V_{(\times)}$ describes the
  simultaneous role of temperature and compactification in the model.

The summand $\Vo$, corresponding to the ``flat''\,model, and the
summand $\VL$, representing the influence of compactification,
were found earlier in eqs. (\ref{www1}) and (\ref{www2});
 in eq. (\ref{4.10}) summand $\Vo$ corresponds to the term which include the first summand (equal to 1) in the first square brackets and the first summand (equal to 1) in the second square brackets,  while summand $\VL$ corresponds to the term which include the first summand in the first square brackets and the second summand in the second square brackets.
As for summands
$V_{(T)}$ and $V_{(\times)}$, which correspond to the second summand in the first square brackets of (\ref{4.10}) and first and second summands in the second square brackets respectively,
we will calculate them
in a different way in
order to show the symmetry between spatial compactification and
usual temporal compactification just related to finite temperature.

Thermal term $V_{(T)}$, i.e. the term that contains the second summand in the first square brackets of (\ref{4.10}) and the first summand in the second square brackets (equal to $1$), can be calculated by integrating over $s$ via the formula
\begin{equation}
\int\limits_{0}^{\infty}x^{-n-1/2}\e^{-px-q/x}dx=(-1)^n\sqrt{\frac{\pi}{p}}\frac{\partial^n}{\partial q^n}\e^{-2\sqrt{pq}}
\end{equation}
with $n=2$ and using the definition of polylogarithm function $\Li_\nu(x)=\sum\limits_{k=1}^{\infty}\frac{x^k}{k^\nu}$. The result of this calculation is similar to that for $\VL$ in eq. (\ref{www2})
\begin{equation}\label{V_T} \begin{gathered}
V_{(T)}= \cfrac{1}{2\pi^{3/2}v_{\rm F}^2} \int\limits_0^\infty
\cfrac{ds}{s^{5/2}} \left[\exp(-s\sigma^2)\right]\times
\left[2\sum\limits_{\ell=1}^{+\infty}(-1)^\ell\exp\left(-\cfrac{\beta^2
\ell^2}{4s}\right)\right]=
\\
=\cfrac{2}{\pi\beta^3 v_{\rm
F}^2}\left[\sigma\beta\Li_2\left(-e^{-\sigma\beta}\right)+\Li_3\left(-e^{-\sigma\beta}\right)\right].
\end{gathered}\end{equation}
The cross-term $V_{(\times)}$ of the effective potential cannot be
expressed in terms of special functions, and we will take it into
account by using numerical calculations
\begin{equation}\begin{gathered}
V_{(\times)}= \cfrac{1}{4\pi^{3/2}v_{\rm F}^2} \int\limits_0^\infty
\cfrac{ds}{s^{5/2}} \left[\exp(-s\sigma^2)\right]\times
\left[2\sum\limits_{\ell=1}^{+\infty}(-1)^\ell\exp\left(-\cfrac{\beta^2
\ell^2}{4s}\right)\right]\times
\\
\times \left[ 2\sum\limits_{n=1}^{+\infty}\exp\left(-\cfrac{L^2
n^2}{4sv_{\rm F}^2}\right)\cos(2\pi n \phi)\right]=
\\
=\cfrac{4}{\pi v_{\rm
F}^2}\sum\limits_{\ell=1}^{+\infty}\sum\limits_{n=1}^{+\infty}
\left\{(-1)^\ell\cos(2\pi n \phi)
\left[\exp\left(-\sigma\sqrt{\beta^2\ell^2+\frac{L^2n^2}{v_{\rm
F}^2}}\right)
\times\cfrac{\sigma\sqrt{\beta^2\ell^2+\frac{L^2n^2}{v_{\rm
F}^2}}+1}{\left(\sqrt{\beta^2\ell^2+\frac{L^2n^2}{v_{\rm
F}^2}}\right)^3} \right]\right\}
.\label{V_cross}\end{gathered}\end{equation}

\begin{figure}[!t]
\begin{center}
\includegraphics [width=170pt]{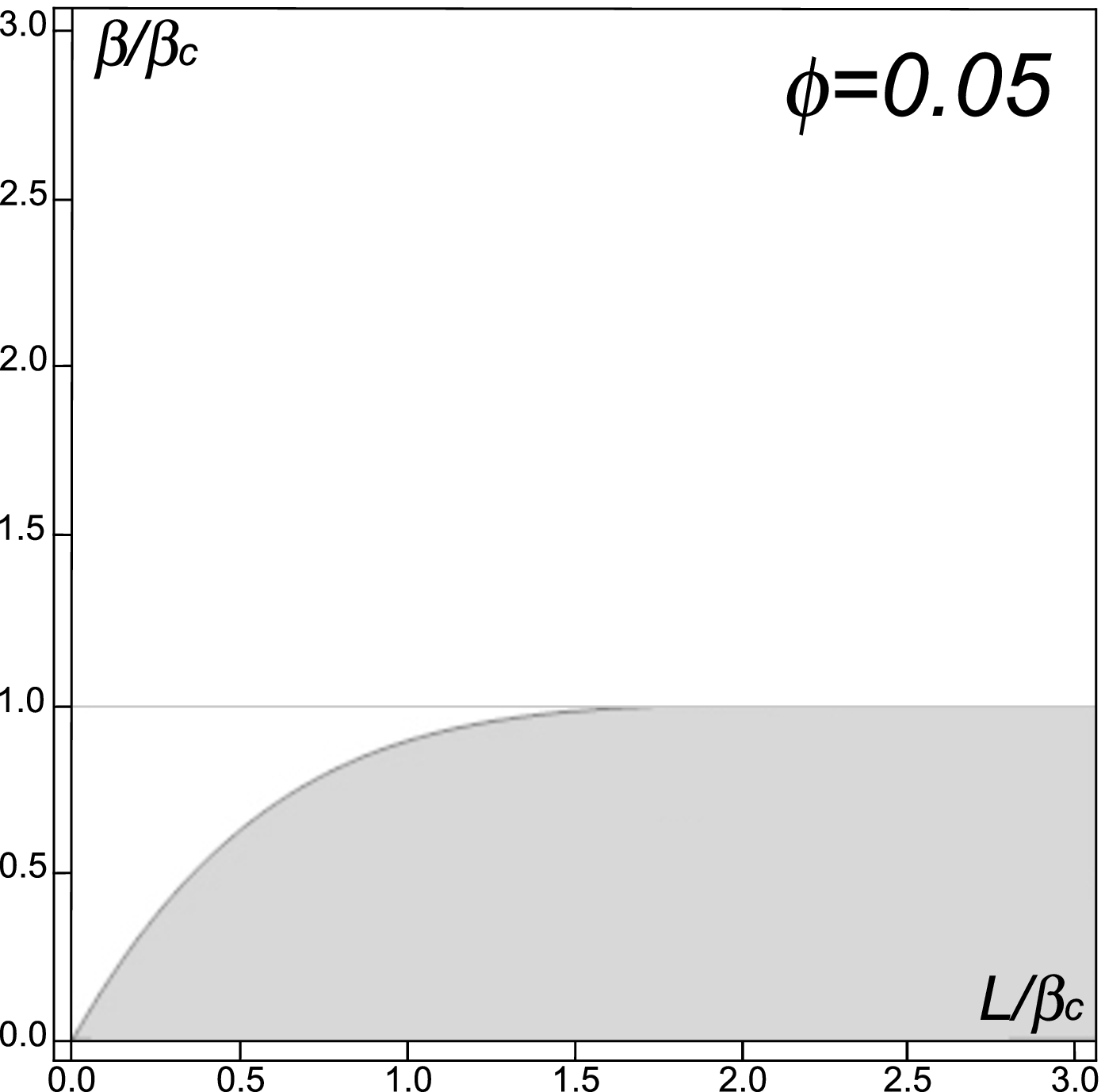}
\includegraphics [width=170pt]{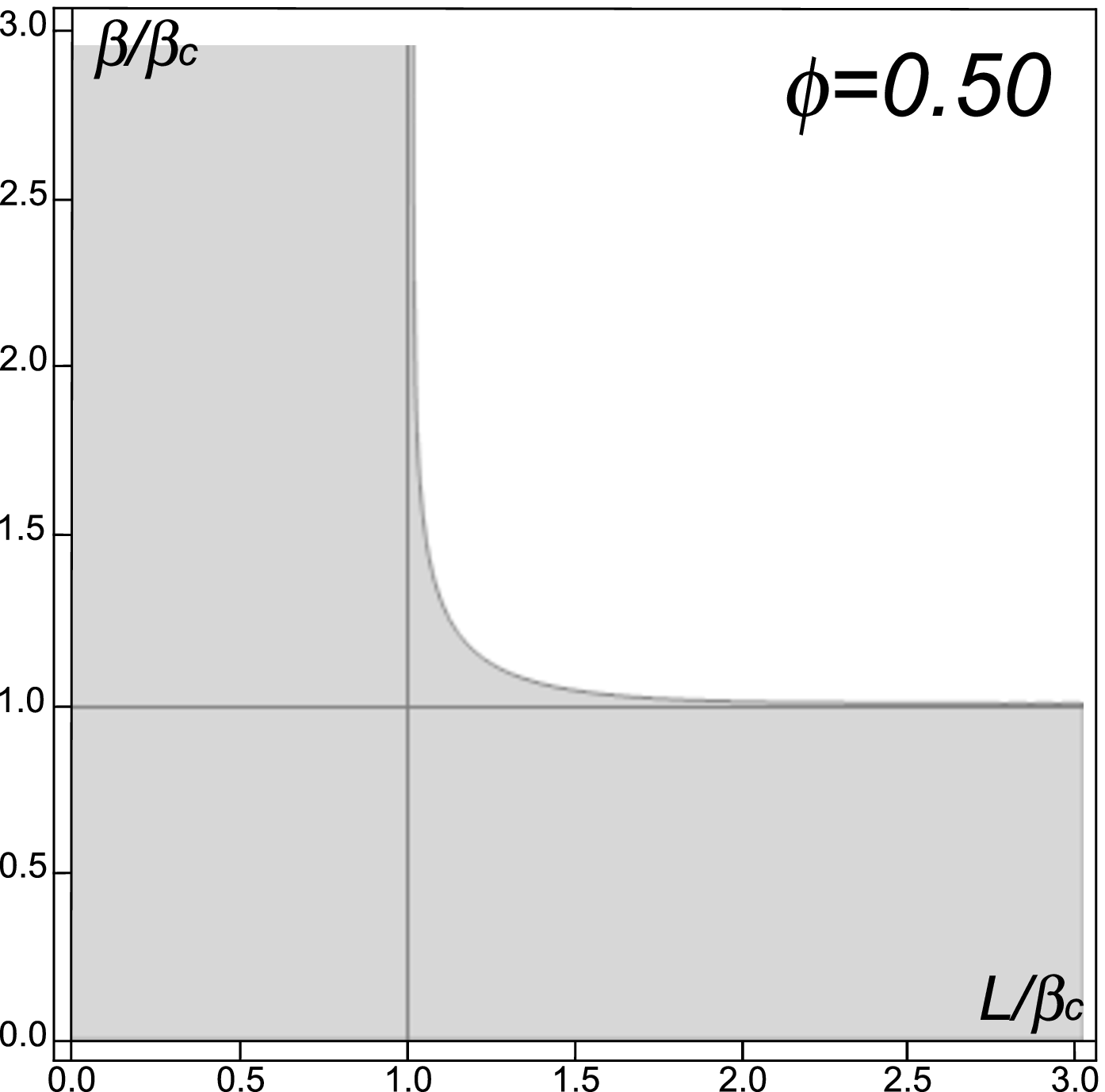}
\\
\includegraphics [width=170pt]{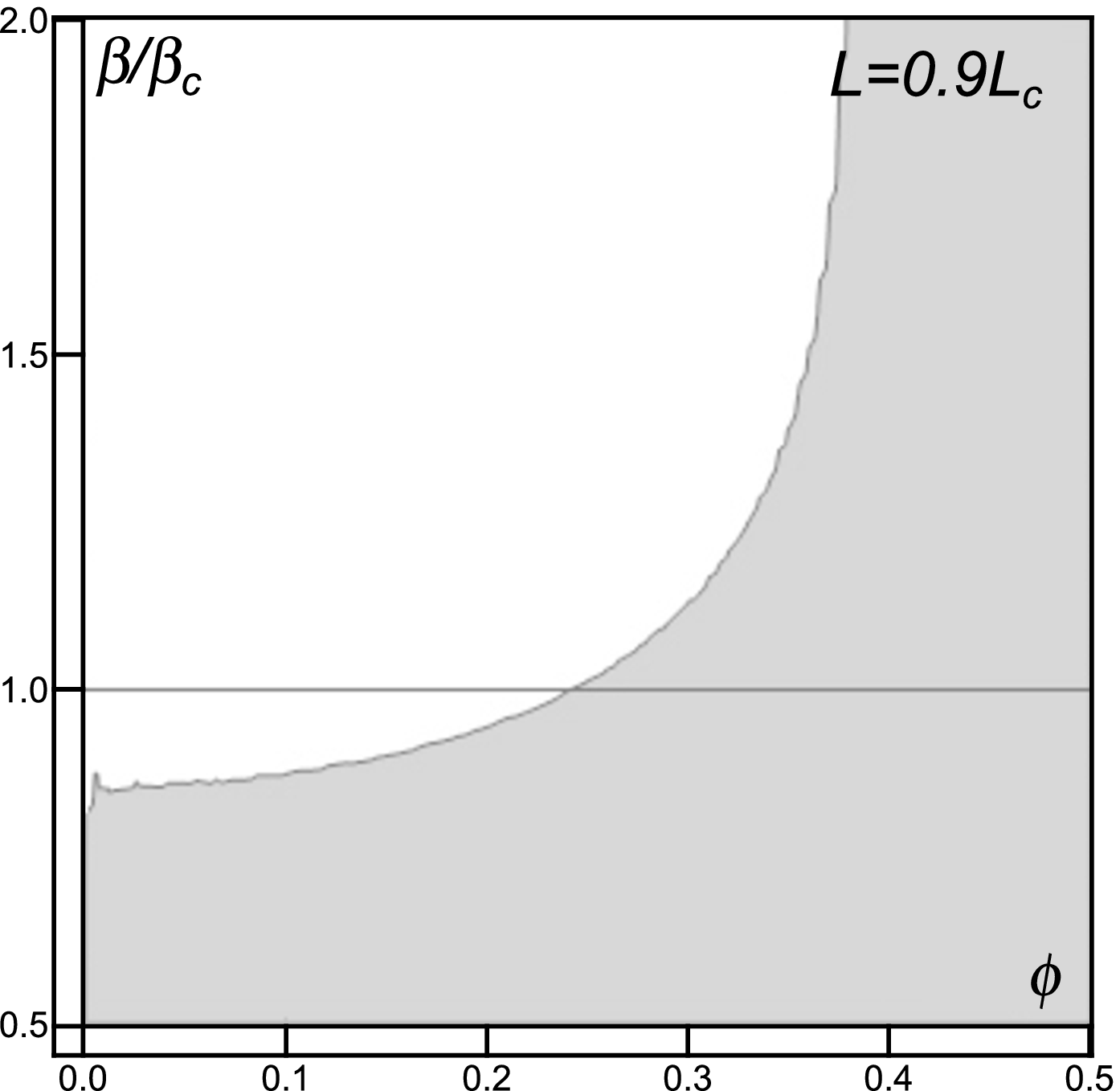}
\end{center}
\caption{\it Phase diagrams of the model in $(L, \beta)$ coordinates
  with different values of
the magnetic phase $\phi$ and in the plane $(\phi,
  \beta)$ with fixed $L<L_c$.The painted area of graphs corresponds to
  the symmetrical phase, while the unpainted one corresponds to the
  broken symmetry. }
\label{fig.AB}
\end{figure}

With the effective potential thus obtained, we can draw
corresponding phase diagrams of the model. Fig. 4 shows phase
diagrams in the $(L,\beta)$ and $(\phi,\beta)$ coordinates. The painted area of graphs corresponds
to the symmetrical phase, while the unpainted one corresponds to the
broken symmetry. Note, that since the constant $g$ is dimensional, we can multiply all dimensional parameters by proper powers of $|g|$ (if $g<0$) to make them dimensionless in the quasi-Plank unit system with $\hbar=c=|g|=1$. Thus, in all the diagrams we assume that $g=-1$.
Evidently, the chiral symmetry is broken in the ``flat'' limit, when $T\to 0, L\to\infty$.
The inverse temperature $\beta$ and compactification length $L$ on
the pictures are normalized by the critical values $\beta_c$ and
$L_c$ introduced earlier. We have chosen $\phi=0.05$ as value close
to zero, since it can be shown that  the phase diagram with
$\phi\equiv 0$ is nearly the same \cite{203}. However, to draw that
diagram one should write the effective potential in a different
form, in which the symmetry between $L$ and $\beta$ is not obvious.
The corresponding  lengthy and straightforward calculation is not
included in this paper. The symmetry between the influence of finite
temperature and spatial compactification mentioned above is
especially evident on the graph corresponding to $\phi=1/2$, which
could be predicted from eq. (\ref{4.10}). The reduction of the
spatial circle length gives thus the same result as an increase of
the temperature. It should be mentioned that symmetry breaking and
dimensional reduction in a 3D GN model were discussed in ref. \cite{202}
with periodic, $\phi=0$, and antiperiodic, $\phi=1/2$, boundary
conditions, but without taking the influence of temperature into
account. We found it interesting that in the case of small $\phi$
spatial compactification counteracts the symmetry restoration caused
by the temperature. It is also interesting that a temperature higher
than the critical one in the ``flat''\,model is necessary to restore
the symmetry, if $\phi$ is small enough. However, the symmetry can
be restored independently of the phase $\phi$ and compactification
length $L$, if the temperature is high enough.

\section{Summary and conclusions}

In this paper we have investigated chiral symmetry breaking in
(2+1)D models with four-fermion interaction that are effectively
used in studying polymers, and especially graphene. In Sect.
\ref{sec2} we started from a ''honeycomb'' graphene-like
lattice-based
Hamiltonian (\ref{2.5}), to which we then added the
four-fermion interaction terms (\ref{2.29}). Via Fierz
transformation (see \ref{AppA}), we then obtained Lagrangian
(\ref{2.34}) and, by omitting any symmetry relations between
coupling constants, finally arrived at the generalized GN-type
schematic model (\ref{2.35}). In Sect. \ref{sec3} we studied the
condensates appearing in the model and found the gap equation
(\ref{3.13}) clearly demonstrating that  the chiral symmetry of the
model can be broken by non-zero condensates, depending on the
magnitude of the coupling constant.

The symmetry properties  of the appearing condensates with respect
to discrete chiral and $\mathcal C$, $\mathcal P$ and $\mathcal T$
transformations are then considered in Table \ref{Tab1}. Some
lengthy calculations and discussions involving the phase structure
of the obtained generalized Gross--Neveu model are given in \ref{X}.
Moreover, in eq. (\ref{3.19}) we have collected the 1PI two-point Green
functions (inverse propagators) of exciton fields resulting from
fermion loop calculations in Appendix C.

In addition, we have also investigated carbon nanotubes with
corresponding boundary conditions (\ref{3.24}) and (\ref{3.25})
incorporating the effect of an external magnetic field. As a result,
we calculated the effective thermodynamic potential for the nanotube
model (\ref{3.35}) including effects of finite temperature, particle
density and external magnetic fields. As illustration, we
numerically investigated in Sect. \ref{sec4} the chiral symmetry
properties after simplifying the model by taking into account only
one condensate. In this case, the model is reduced to a standard
Gross--Neveu model with only a discrete $\gamma^5$ chiral symmetry.
The phase diagram in the $(\mu,T)$ plane is drawn in
Fig.~\ref{Fig.muT} which shows its great similarity with the phase
diagram of the 2D Gross--Neveu model presented in ref. \cite{n2c}.

Finally, we investigated the phase structure under the influence of
the Aharonov--Bohm effect and showed that this effect can greatly
influence the phase structure of the system (Fig.~\ref{fig.AB}).
Here we can notice that, depending on the Aharonov--Bohm phase,
compactification of spatial dimension can either oppose or assist
the thermal restoration of the originally broken chiral symmetry,
i.e. in the case of periodical boundary conditions restoration of
chiral symmetry requires a greater temperature than in the planar
model. On the contrary, with an antiperiodic boundary condition
(that can be provided by the Aharonov--Bohm effect), the temperature
required for the symmetry restoration is lower than in the planar
model. Then, if the radius of the compactified dimension is small
enough, the chiral symmetry can be restored even at zero
temperature.

\section*{Acknowledgments}

One of the authors (D.E.) is grateful to G.W. Semenoff, H.Reinhardt, V.V. Braguta, and V.A. Miransky for useful discussions.

\appendix

\section{Fierz-transformed interactions \label{AppA}}

For completeness, we compile in this appendix the formulas, required
to obtain the Fierz-transformed expressions of the four-fermion
interaction in eqs. (\ref{2.32}) and (\ref{2.33}) of the text.

Let us consider the 16 Hermitian $4\times4$ matrices of the Dirac
algebra $\left\{\Gamma\right\}^{16}_{A=1}$ quoted in
eq. (\ref{2.30}). Then, any Hermitian matrix M can be written as
$M=\frac{1}{4}\left({\rm Tr} M \Gamma^A\right)\Gamma^A$, where
summation over repeated indices is understood. The last expression
for $M$ can be rewritten as
\begin{equation}
4\delta_{\ell i}\delta_{m j}M_{\ell
m}=\Gamma^{A}_{m\ell}\Gamma^A_{ij}M_{\ell m} ,
\label{A.1}\end{equation} leading to the identity $\delta_{\ell
i}\delta_{mj}=\frac{1}{4}\Gamma^A_{m\ell}\Gamma^A_{ij}$. Using this
identity, one can rewrite the product of two matrix elements as
\begin{equation}
M_{ij}N_{mn}=\frac{1}{16}\left({\rm Tr} M\Gamma^A N
\Gamma^B\right)\Gamma^B_{in}\Gamma^A_{mj} .
\label{A.2}\end{equation} This then leads to the required expansion
of a four-fermion term as \cite{n24}
\begin{equation} \begin{gathered}
\left[\psibar^{(a,s)}(x)M\psi^{(b,s')}(x)\right]\left[\psibar^{(c,s'')}(y)N\psi^{(d,s''')}(y)\right]
\left(\delta^{ab}\delta^{cd}\right)\left(\delta^{ss'}\delta^{s''s'''}\right)
\\
=-\frac{1}{16}\left({\rm Tr}M\Gamma^A N \Gamma^B\right)
\left[\psibar^{(a,s)}(x)\Gamma^B \psi^{(d,s''')}(y)
\right]\left[\psibar^{(c,s'')}(y)\Gamma^A \psi^{(b,s')}(x)\right]
\left(\delta^{ab}\delta^{cd}\right)\left(\delta^{ss'}\delta^{s''s'''}\right)
\label{A.3}\end{gathered}\end{equation}
which is used in the text
for $x=y$. The minus sign in the last line arises from the Grassmann
nature of the fermion fields. In a second step, we have to
Fierz-transform the spin and flavor singlet structure of the
interaction terms by using the completeness relations for
the spin and flavor groups $U(S)$,  $U(N)$,
\begin{equation}
\frac{1}{2}\delta^{ss'}\delta^{s''s'''}=\frac{1}{4}\delta^{ss'''}\delta^{s''s'}+
\sum\limits_{m=1}^{3}\left(\frac{\sigma^m}{2}\right)^{ss'''}\left(\frac{\sigma^m}{2}\right)^{s''s'}
,\label{A.4}\end{equation}
\begin{equation}
\frac{1}{2}\delta_{mk}\delta_{i\ell}=\frac{1}{2N}\delta_{m\ell}\delta_{ik}+
\sum\limits_{\alpha=1}^{N^2-1}\left(\frac{\lambda^\alpha}{2}\right)_{m\ell}
\left(\frac{\lambda^\alpha}{2}\right)_{ik}
,\label{A.5}\end{equation}
\begin{equation} \nonumber
{\rm
tr}\frac{\lambda^\alpha}{2}\frac{\lambda^\beta}{2}=\frac{1}{2}\delta^{\alpha\beta}\;\text{etc.}
\end{equation}

Since we will only consider condensates of flavor/spin singlet
excitons, we shall keep here only the first terms in the r.h.s. of
eqs. (\ref{A.4}) and (\ref{A.5}) and discard the other ones.

Expression (\ref{A.3}) then simplifies to the form
\begin{equation}
\left[\psibar M\psi\right]\left[\psibar N \psi\right]
=-\frac{1}{16N_{\rm f}}\left({\rm Tr}M\Gamma^A N \Gamma^B\right)
\left[\psibar \Gamma^B\psi\right]\left[\psibar \Gamma^A \psi\right]
,\label{A.6}\end{equation} with $N_{\rm f}=2N$, $\left[\psibar
M\psi\right]=\psibar^{as} M\psi^{as}$ etc., and a summation over
repeated spin/flavor indices $(s,a)$ is understood. Obviously, in
our case we have $M=N$, and we have to apply eq.(\ref{A.3}) for
$M\otimes M=\left\{\gamma^0\otimes\gamma^0, t^0\otimes t^0\right\}$.
In order to quote the required Fierz-transformed expressions, it is
convenient to introduce the following notations:
\begin{equation} \begin{gathered}
\overrightarrow{V}=\left\{\psibar\Gamma_i\psi\right\},
~~~~\Gamma_i=\{{\rm
    I}_4,\gamma^3,i\gamma^5\},
~~~~S=\psibar\gamma^{35}\psi,
\\
\overrightarrow{V}^\mu=\left\{\psibar\gamma^\mu\Gamma_i^*\psi\right\},
~~~~\Gamma_i^*=\{{\rm I}_4,i\gamma^3,\gamma^5\},~~~~
S^\mu=\psibar\gamma^\mu\gamma^{35}\psi, \label{A.7}
\end{gathered}\end{equation}
(Note that $\Gamma_i^*$ is not the conjugate of $\Gamma_i$.) For
comparison, let us start with the Lorentz- and chiral invariant
Thirring-like four-fermion term
\begin{equation}
\left(\psibar\gamma_\mu\psi\right)\left(\psibar\gamma^\mu\psi\right)=
-\frac{1}{4N_{\rm f}} \left\{
3\left(\overrightarrow{V}^2+S^2\right)-\left(\overrightarrow{V}_\mu\cdot\overrightarrow{V}^\mu+S_\mu
S^\mu\right) \right\} . \label{A.8}\end{equation} The Lorentz
non-invariant, but chiral invariant Coulomb-type interaction and the
chiral symmetry breaking Lorentz-invariant terms, considered in the
text, transform as follows
\begin{equation}
\left[\psibar\gamma^0\psi\right]\left[\psibar\gamma^0\psi\right]=
-\frac{1}{4N_{\rm f}} \left\{
\left(\overrightarrow{V}^2+S^2\right)+\overrightarrow{V}^\mu\cdot\overrightarrow{V}^\mu+S^\mu
S^\mu \right\} . \label{A.9}\end{equation}
\begin{equation}
\left[\psibar\psi\right]\left[\psibar\psi\right]= -\frac{1}{4N_{\rm f}}
\left\{
\left({\overrightarrow{V}^*}^{2}+S^2\right)+{\overrightarrow{V}^*}_\mu\cdot{\overrightarrow{V}^*}^\mu+S_\mu
S^\mu \right\} . \label{A.10}\end{equation} Here we used the
notations
\begin{equation} \begin{gathered}
\overrightarrow{V}_\mu\cdot\overrightarrow{V}^\mu=
g_{\mu\nu}\overrightarrow{V}^\mu\cdot\overrightarrow{V}^\nu=
\overrightarrow{V}^0\cdot\overrightarrow{V}^0-\overrightarrow{V}^i\cdot\overrightarrow{V}^i,
\\
\overrightarrow{V}^\mu\cdot\overrightarrow{V}^\mu=
\overrightarrow{V}^0\cdot\overrightarrow{V}^0+\overrightarrow{V}^i\cdot\overrightarrow{V}^i,
{\rm etc}.,
\\
\label{A.11}
\overrightarrow{V}^*=\left\{\psibar\Gamma^*_i\psi\right\}, ~~~~
{\overrightarrow{V}^*}^\mu=\left\{\psibar\gamma^\mu\Gamma_i\psi\right\}
.\end{gathered}\end{equation} In the text we have only considered
the NJL-type scalar/pseudoscalar interactions of flavor/spin-singlet
type and discarded, for simplicity, non-singlet axial/vector type
interactions.

\section{Phase structure of the generalized Gross--Neveu model (\ref{2.35})}\label{X}

The phase structure of the schematic model (\ref{2.35}) is described
by the effective potential (\ref{3.9}), where, for brevity of
notations, we put here $v_{\rm F}=1$, and, for convenience, shift also the
absolute value sign in the original definition of $M_k$ to the term
$|M_k|^3$
\begin{eqnarray}
V(\sigma_i, \varphi_i)=\sum_{k=1}^2\left
[\frac{g_k}{4}\sigma_k^2+\frac{h_k}{4}\varphi_k^2+\frac{|M_k|^3}{6\pi}\right
]. \label{X1}
\end{eqnarray}
Thus we now have $M_{1}=\sigma_2
+\sqrt{\sigma_1^2+\varphi_1^2+\varphi_2^2}$, $M_{2}=\sigma_2
-\sqrt{\sigma_1^2+\varphi_1^2+\varphi_2^2}$.  Since the function
(\ref{X1}) is even with respect to each variable $\sigma_{1,2}$ and
$\varphi_{1,2}$, i.e. it is invariant under each of the
transformations $\sigma_1\to-\sigma_1$, $\sigma_2\to-\sigma_2$,
$\varphi_1\to-\varphi_1$, and $\varphi_2\to-\varphi_2$, we can
suppose that in eq. (\ref{X1}) $\varphi_{1,2}\ge 0$ and $\sigma_{1,2}\ge
0$. Our goal is to find the global minimum point (GMP) of the
effective potential (\ref{X1}) vs $\varphi_{1,2}\ge 0$ and
$\sigma_{1,2}\ge 0$. However, the structure of the function
(\ref{X1}) tells us to use at the beginning another set of
independent variables. Namely, it is convenient first to study its
extremal properties in terms of $M_1$, $M_2$, $x$ and $y$, where
$x=\varphi_1^2$, $y=\varphi_2^2$, and then return to the original
variables $\varphi_{1,2}\ge 0$ and $\sigma_{1,2}\ge 0$. Since
$\sigma_1^2=-x-y+(M_1-M_2)^2/4$ and $\sigma_2^2=(M_1+M_2)^2/4$, we
have instead of eq. (\ref{X1}) the following function
\begin{eqnarray}
V(M_1,M_2,x,y)=\frac{g_1}{16}(M_1-M_2)^2+\frac{g_2}{16}(M_1+M_2)^2+\frac{h_1-g_1}{4}x+\frac{h_2-g_1}{4}y+\frac{|M_1|^3}{6\pi}+\frac{|M_2|^3}{6\pi}.
\label{X2}
\end{eqnarray}
Note, there are natural restrictions on the new variables, $M_1\ge
0$, $-\infty <M_2<\infty$, $x,y\ge 0$ and $x+y\le (M_1-M_2)^2/4$. In
order to find the GMP of the function (\ref{X2}), we use the
following strategy. First, we will minimize it (at fixed $M_{1,2}$)
with respect to $x$ and $y$ by varying in the compact and closed
domain $\omega =\{(x,y):x,y\ge 0,~x+y\le (M_1-M_2)^2/4\}$. Second,
the obtained minimal expression of the effective potential will then
be minimized over $M_{1,2}$. Since $V(M_1,M_2,x,y)$ is a linear
function in both $x$ and $y$, it is obvious that its least value on
the triangle region $\omega$ is reached in one of the vertices of
this triangle, i.e. in one of the points $(x_1=0,y_1=0)$,
$(x_2=(M_1-M_2)^2/4,y_2=0)$ and $(x_3=0, y_3=(M_1-M_2)^2/4)$. There,
the effective potential (\ref{X2}) takes the following values
\begin{eqnarray}
V_I(M_1,M_2)&\equiv& V(M_1,M_2,x=0,y=0)=\frac{g_1}{16}(M_1-M_2)^2+\frac{g_2}{16}(M_1+M_2)^2+\frac{M_1^3}{6\pi}+\frac{|M_2|^3}{6\pi}, \label{X3}\\
V_{II}(M_1,M_2)&\equiv& V\left (M_1,M_2,x=\frac{(M_1-M_2)^2}4,y=0\right )=\frac{h_1}{16}(M_1-M_2)^2+\frac{g_2}{16}(M_1+M_2)^2+\frac{M_1^3}{6\pi}+\frac{|M_2|^3}{6\pi}, \label{X4}\\
V_{III}(M_1,M_2)&\equiv& V\left
(M_1,M_2,x=0,y=\frac{(M_1-M_2)^2}4\right
)=\frac{h_2}{16}(M_1-M_2)^2+\frac{g_2}{16}(M_1+M_2)^2+\frac{M_1^3}{6\pi}+\frac{|M_2|^3}{6\pi}.
\label{X5}
\end{eqnarray}
To compare the quantities (\ref{X3})-(\ref{X5}), let us fix the
value of the coupling constant $g_1$ and divide the plane of the
couplings $h_1$ and $h_2$ into three regions $I$, $II$ and $III$
(see Fig. 5 for the case $g_1>0$), where $I=\{(h_1,h_2):~h_1>g_1,
h_2>g_1\}$, $II=\{(h_1,h_2):~ h_1<g_1, h_2>h_1\}$ and
$III=\{(h_1,h_2):~ h_2<g_1, h_2<h_1\}$. Then a direct comparison of
the functions (\ref{X3})-(\ref{X5}) shows that (i) in the region $I$
the GMP of the effective potential (\ref{X2}) with respect to $x$
and $y$ is the point $(x_1=0,y_1=0)$, where the least value of
$V(M_1,M_2,x,y)$ is equal to the function $V_I(M_1,M_2)$. (ii) If
the couplings $h_1$ and $h_2$ are in the region $II$, then the least
value of the effective potential (\ref{X2}) vs $x$ and $y$ is
reached at the point $(x_2=(M_1-M_2)^2/4,y_2=0)$, where it is the
quantity $V_{II}(M_1,M_2)$. (iii) Finally, if $(h_1,h_2)\in III$,
then the GMP of the function (\ref{X2}) over the variables $x$ and
$y$ is realized at the point $(x_3=0, y_3=(M_1-M_2)^2/4)$ and the
least value of (\ref{X2}) is the quantity  $V_{III}(M_1,M_2)$. Now,
we will find the GMPs of the functions $V_{I}(M_1,M_2)$,
$V_{II}(M_1,M_2)$ and $V_{III}(M_1,M_2)$ vs $M_1$ and $M_2$ from the
region  $M_1\ge 0$, $-\infty <M_2<\infty$.
\begin{figure}
\includegraphics[width=0.45\textwidth]{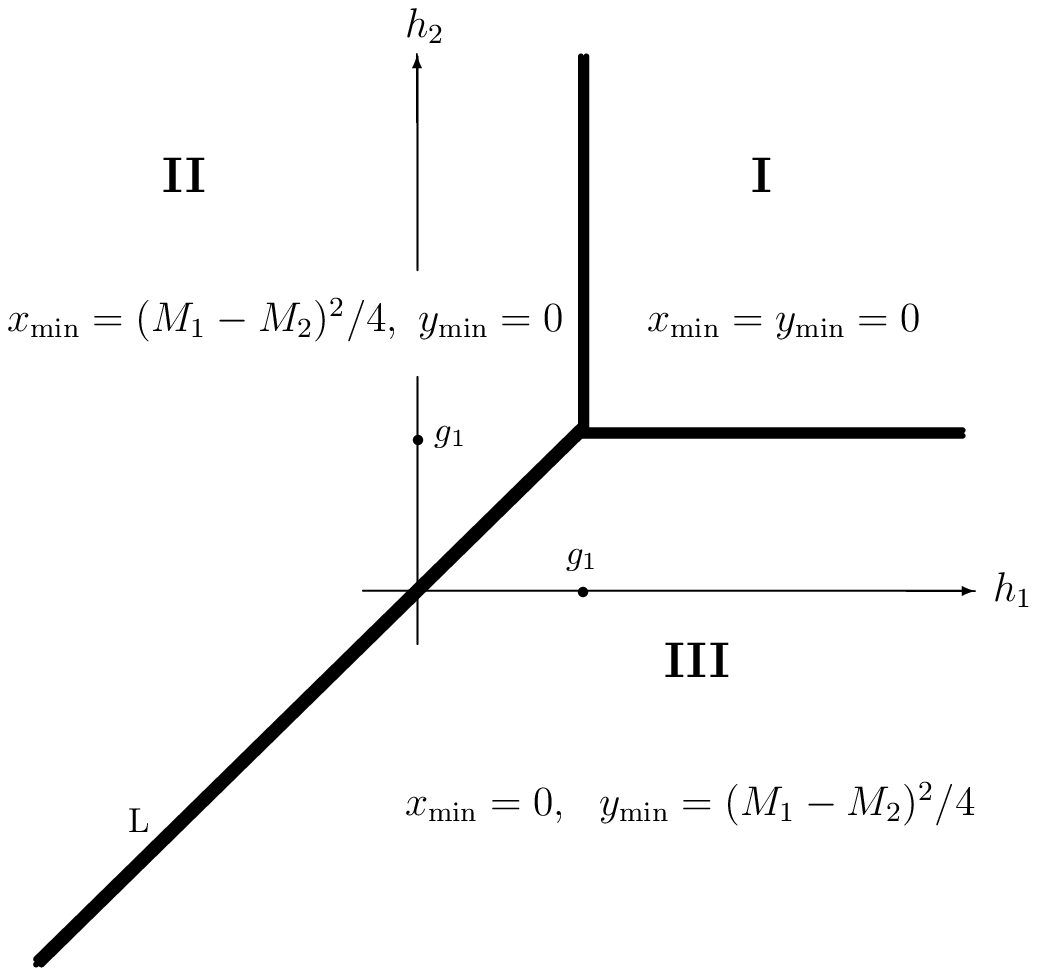}
\hfill
\includegraphics[width=0.43\textwidth]{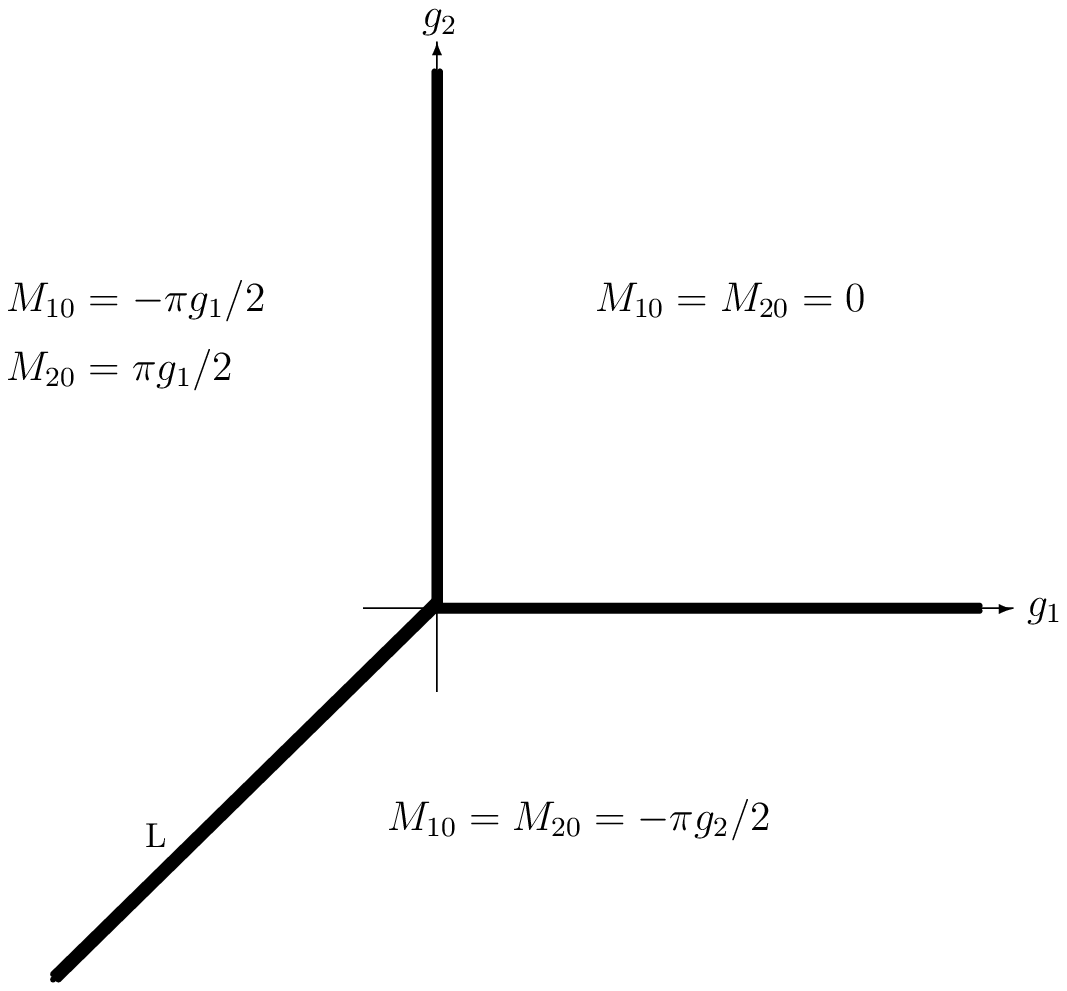}\\
\parbox[t]{0.45\textwidth}{
\caption{\it The plane of coupling constants $h_1$ and $h_2$ is
divided
  into three regions I, II and III. In each region the coordinates
  $x_{\rm min}$ and $y_{\rm min}$ of the least value point of the
  function (\ref{X2}) vs $(x,y)\in\omega =\{(x,y):x,y\ge 0,~x+y\le
  (M_1-M_2)^2/4\}$
are presented.  The line L is defined by the relation  L
$\equiv\{(h_1,h_2):h_1=h_2\}$. For simplicity, the coupling constant
$g_1$ is selected to be positive
and determines here the origin $(g_1,g_1)$ of the thick axis system.
 }}\hfill
\parbox[t]{0.45\textwidth}{
\caption{\it The coordinates $M_{10}$ and $M_{20}$ of the global
minimum point of the function $V_I(M_1,M_2)$ (\ref{X3}) in
dependence on the coupling constants $g_1$, $g_2$. The line L is
defined by the relation  L $\equiv\{(g_1,g_2):g_1=g_2\}$.} }
\end{figure}

We start from the case, when $(h_1,h_2)\in I$, i.e. from finding of
a GMP of the function $V_{I}(M_1,M_2)$. There is a system of two
stationarity equations,
\begin{eqnarray}
\frac{\partial V_I(M_1,M_2)}{\partial M_1}&\equiv& \frac{g_1}{8}(M_1-M_2)+\frac{g_2}{8}(M_1+M_2)+\frac{M_1^2}{2\pi}=0~,\nonumber \\
\frac{\partial V_I(M_1,M_2)}{\partial M_2}&\equiv&
\frac{g_1}{8}(M_2-M_1)+\frac{g_2}{8}(M_1+M_2)+{\rm
sign}(M_2)\frac{M_2^2}{2\pi}=0~,\label{X6}
\end{eqnarray}
where ${\rm sign}(x)$ is the sign function. The GMP
$(M_{10},M_{20})$ of the function $V_{I}(M_1,M_2)$ is a solution of
the system of stationarity equations (\ref{X6}). Moreover, it
depends on the values of the coupling constants $g_1$ and $g_2$.
Solving the system (\ref{X6}), one can find the behavior of the
coordinates $M_{10}$ and $M_{20}$ vs $g_1$ and $g_2$ (see Fig. 6,
where the plane $(g_1,g_2)$ is divided into three regions
corresponding to different expressions for $M_{10}$ and $M_{20}$).
Namely, if $g_{1,2}>0$, then $M_{10}=M_{20}=0$. If $g_2<0$ and
$g_2<g_1$, then $M_{10}=M_{20}=-\pi g_2/2$. If $g_1<0$ and
$g_2>g_1$, then $M_{10}=-M_{20}=-\pi g_1/2$. Hence, returning to the
original variables $\varphi_{1,2}\ge 0$ and $\sigma_{1,2}\ge 0$, one
can establish the GMP of the effective potential (\ref{X1}) and, as
a result, the form of the ground state expectation values
$\langle\sigma_{1,2}\rangle$, $\langle\varphi_{1,2}\rangle$ when
$(h_1,h_2)\in I$. So we find that if
\begin{eqnarray}
(g_1,g_2,h_1,h_2)\in\Omega_{I0}\equiv\{(g_1,g_2,h_1,h_2):~h_1>g_1,
h_2>g_1;g_{1}>0,g_{2}>0\}, \label{70}
\end{eqnarray}
then all ground state expectation values are zero,
$\langle\sigma_{1,2}\rangle=0$ and $\langle\varphi_{1,2}\rangle=0$.
If
\begin{eqnarray}
(g_1,g_2,h_1,h_2)\in\Omega_{I\sigma_1}\equiv\{(g_1,g_2,h_1,h_2):~h_1>g_1,
h_2>g_1;g_{1}<0,g_{2}>g_1\},\label{80}
\end{eqnarray}
then $\langle\sigma_{1}\rangle=-\pi g_1/2$,
$\langle\sigma_{2}\rangle=0$, $\langle\varphi_{1,2}\rangle=0$.
Finally, if
\begin{eqnarray}
(g_1,g_2,h_1,h_2)\in\Omega_{I\sigma_2}\equiv\{(g_1,g_2,h_1,h_2):~h_1>g_1,
h_2>g_1;g_{2}<0,g_{2}<g_1\},\label{90}
\end{eqnarray}
then $\langle\sigma_{2}\rangle=-\pi g_2/2$,
$\langle\sigma_{1}\rangle=0$, $\langle\varphi_{1,2}\rangle=0$.

If the point $(h_1,h_2)$ belongs to the region $II$, then we need to
study the extrema properties of the function $V_{II}(M_1,M_2)$.
Since this function is obtained from (\ref{X3}) by the replacement
$g_1\to h_1$, the behavior of its GMP vs $h_1$ and $g_2$ can be
easily found from Fig. 6 also by an evident replacement $g_1\to
h_1$. After that it is possible to get the form of ground state
expectation values of the fields $\varphi_{1,2}$ and $\sigma_{1,2}$.
Namely, if
\begin{eqnarray}
(g_1,g_2,h_1,h_2)\in\Omega_{II0}\equiv\{(g_1,g_2,h_1,h_2):~h_1<g_1,
h_2>h_1;h_{1}>0,g_{2}>0\}, \label{7}
\end{eqnarray}
then $\langle\sigma_{1,2}\rangle=0$ and
$\langle\varphi_{1,2}\rangle=0$. If
\begin{eqnarray}
(g_1,g_2,h_1,h_2)\in\Omega_{II\varphi_1}\equiv\{(g_1,g_2,h_1,h_2):~h_1<g_1,
h_2>h_1;h_{1}<0,g_{2}>h_1\}, \label{8}
\end{eqnarray}
we have $\langle\sigma_{1}\rangle=\langle\sigma_{2}\rangle=0$,
$\langle\varphi_{1}\rangle=-\pi h_1/2$,
$\langle\varphi_{2}\rangle=0$. If
\begin{eqnarray}
(g_1,g_2,h_1,h_2)\in\Omega_{II\sigma_2}\equiv\{(g_1,g_2,h_1,h_2):~h_1<g_1,
h_2>h_1;g_{2}<0,g_{2}<h_1\}, \label{9}
\end{eqnarray}
then $\langle\sigma_{2}\rangle=-\pi g_2/2$,
$\langle\sigma_{1}\rangle=0$, $\langle\varphi_{1,2}\rangle=0$. In a
similar way it is easy to establish the structure of the condensates
$\langle\sigma_{1,2}\rangle$ and $\langle\varphi_{1,2}\rangle$ in
the case when the point $(h_1,h_2)$ belongs to the region $III$. So,
we see that if
\begin{eqnarray}
(g_1,g_2,h_1,h_2)\in\Omega_{III0}\equiv\{(g_1,g_2,h_1,h_2):~h_2<g_1,
h_2<h_1;h_{2}>0,g_{2}>0\}, \label{10}
\end{eqnarray}
then $\langle\sigma_{1,2}\rangle=0$ and
$\langle\varphi_{1,2}\rangle=0$. If
\begin{eqnarray}
(g_1,g_2,h_1,h_2)\in\Omega_{III\varphi_2}\equiv\{(g_1,g_2,h_1,h_2):~h_2<g_1,
h_2<h_1;h_{2}<0,g_{2}>h_2\}, \label{11}
\end{eqnarray}
we have $\langle\sigma_{1}\rangle=\langle\sigma_{2}\rangle=0$,
$\langle\varphi_{2}\rangle=-\pi h_2/2$,
$\langle\varphi_{1}\rangle=0$. Finally, if
\begin{eqnarray}
(g_1,g_2,h_1,h_2)\in\Omega_{III\sigma_2}\equiv\{(g_1,g_2,h_1,h_2):~h_2<g_1,
h_2<h_1;g_{2}<0,g_{2}<h_2\}, \label{12}
\end{eqnarray}
then $\langle\sigma_{2}\rangle=-\pi g_2/2$,
$\langle\sigma_{1}\rangle=0$, $\langle\varphi_{1,2}\rangle=0$.

We summarize the results of our investigation of the effective
potential (\ref{X1}) in the Table \ref{T1}. (Note that in order to
apply the data of this table to the model (\ref{2.35}) and/or to the
effective potential (\ref{3.9}), one should  perform the
replacements $g_i\to g_iv_{\rm F}$ and $h_i\to h_iv_{\rm F}$ there.) Each nontrivial
combination of condensates $\langle\sigma_{1,2}\rangle$ and
$\langle\varphi_{1,2}\rangle$, listed in  Table \ref{T1}, corresponds
to some broken discrete symmetries from the set $\{\mathcal P,
\mathcal C, \mathcal T, \gamma^5, \gamma^3\}$, i.e. to some
nontrivial phase of the model. The list of all possible phases of
the model (\ref{2.35}), including the trivial phase with
$\langle\sigma_{1,2}\rangle=0$ and $\langle\varphi_{1,2}\rangle=0$,
can be easily established with the help of  Tables \ref{Tab1} and \ref{T1} (see
also the symmetry properties of the condensates in Sect.~2.4).
\renewcommand{\arraystretch}{0.9}
\renewcommand{\tabcolsep}{0.5cm}
\begin{table*}[!t]
\centering
\begin{tabular}{|c | c c c c|}\hline\hline
$(g_1,g_2,h_1,h_2)$      &  $\langle\sigma_1\rangle$   &
$\langle\sigma_2\rangle$   &   $\langle\varphi_1\rangle$   &
$\langle\varphi_2\rangle$   \\ \hline
$(g_1,g_2,h_1,h_2)\in\Omega_{I0}\cup\Omega_{II0}\cup\Omega_{III0}$   &    0   &   0   &   0   &   0  \\
$(g_1,g_2,h_1,h_2)\in\Omega_{I\sigma_1}$  &   $-\pi g_1/2$   &  0    &   0   &  0  \\
$(g_1,g_2,h_1,h_2)\in\Omega_{I\sigma_2}\cup\Omega_{II\sigma_2}\cup\Omega_{III\sigma_2}$  & 0  &  $-\pi g_2/2$   &   0   &  0  \\
$(g_1,g_2,h_1,h_2)\in\Omega_{II\varphi_1}$  & 0   & 0   &   $-\pi h_1/2$   &  0  \\
$(g_1,g_2,h_1,h_2)\in\Omega_{III\varphi_2}$  &  0  & 0  & 0   &   $-\pi h_2/2$   \\
 \hline\hline
\end{tabular}
\caption{\it\label{T1} The values of the condensates
$\langle\sigma_1\rangle$, $\langle\sigma_2\rangle$,
$\langle\varphi_1\rangle$ and $\langle\varphi_2\rangle$   in
dependence on the coupling constants $g_1$, $g_2$, $h_1$ and $h_2$.
The regions $\Omega_{I0}$,...,$\Omega_{III\varphi_2}$ are defined in
(\ref{70}),..., (\ref{12}), correspondingly.}
\end{table*}

\section{Calculations of the 1PI Green functions (\ref{3.19})}\label{Y}

Choosing e.g. $\hat t_k={\rm I}_4$ in eq. (\ref{3.17}), we have for the 1PI 2-point Green function $\Gamma_{\sigma_1\sigma_1}(x-y)$ of $\sigma_1(x)$ and $\sigma_1(y)$ fields the following expression
\begin{eqnarray}
\Gamma_{\sigma_1\sigma_1}(z)\equiv\frac{\delta^2S^{(2)}_{\rm eff}}{\delta\sigma_1(x)\delta\sigma_1(y)}
=-\frac{1}{2v_{\rm F}G_{1}}\delta^{(3)}(z)+i{\rm
Tr}_{s}\left [G_0(z)G_0(-z)\right ], \label{Y1}
\end{eqnarray}
where  $z=x-y$ and the matrix elements of the propagator $G_0(z)$ are presented in eq. (\ref{3.18}). Introducing the momentum-space representation $\Gamma_{\sigma_1\sigma_1}(p)$ for the Green function (\ref{Y1})
\begin{eqnarray}
\Gamma_{\sigma_1\sigma_1}(p)= \int d^3z \Gamma_{\sigma_1\sigma_1}(z)e^{ipz},\label{Y2}
\end{eqnarray}
and applying the corresponding Fourier transformation to both sides of eq. (\ref{Y1}), using there the expression (\ref{3.18}), one can find
\begin{eqnarray}
\Gamma_{\sigma_1\sigma_1}(p)=-\frac{1}{2v_{\rm F}G_{1}}+i \int\frac{d^3k}{(2\pi)^3}{\rm
Tr}_{s}\left [\left
(\frac{1}{\lefteqn{/}\widetilde k+\lefteqn{/}\widetilde p-\langle\sigma_1\rangle }\right
)\left
(\frac{1}{\lefteqn{/}\widetilde k-\langle\sigma_1\rangle }\right
)\right ].\label{Y4}
\end{eqnarray}
It is clear from the stationarity equations for the effective potential (\ref{3.5}) that
\begin{eqnarray}
\frac{1}{2v_{\rm F}G_{1}}=i\int\frac{d^3k}{(2\pi)^3}\frac{4}{\widetilde k^2-\langle\sigma_1\rangle^2}
\label{Y5}
\end{eqnarray}
in the case $\langle\sigma_1\rangle\ne 0$, $\langle\sigma_2\rangle = 0$, $\langle\varphi_1\rangle = 0$ and $\langle\varphi_2\rangle = 0$.
 Using eq. (\ref{Y5}) in the expression (\ref{Y4}), we have after trace calculation:
\begin{eqnarray}
\Gamma_{\sigma_1\sigma_1}(p)
&=&i \int\frac{d^3k}{(2\pi)^3}\left [\frac{-4\widetilde p^2-4\widetilde k\widetilde p+8\langle\sigma_1\rangle^2}{(\widetilde k^2-\langle\sigma_1\rangle^2)((\widetilde k+\widetilde p)^2-\langle\sigma_1\rangle^2)}\right ].\label{Y7}
\end{eqnarray}
Let us apply in eq. (\ref{Y7}) the general relation
\begin{eqnarray}
\frac 1{AB}=\int_0^1d\alpha\frac{1}{[A\alpha+ B(1-\alpha)]^2}.
\label{Y8}
\end{eqnarray}
Then we have:
\begin{eqnarray}
\Gamma_{\sigma_1\sigma_1}(p)&=&i\int_0^1d\alpha \int\frac{d^3k}{(2\pi)^3}\frac{-4\widetilde p^2-4\widetilde k\widetilde p+8\langle\sigma_1\rangle^2}{[(\widetilde k^2-\langle\sigma_1\rangle^2)\alpha+((\widetilde k+\widetilde p)^2-\langle\sigma_1\rangle^2)(1-\alpha)]^2}\nonumber\\
&=&i\int_0^1d\alpha \int\frac{d^3k}{(2\pi)^3}\frac{-4\widetilde p^2-4\widetilde k\widetilde p+8\langle\sigma_1\rangle^2}{[\widetilde p^2\alpha (1-\alpha)+(\widetilde k+\alpha\widetilde p)^2-\langle\sigma_1\rangle^2]^2}.\label{Y9}
\end{eqnarray}
Now let us change variables in the $k$-integration  in eq. (\ref{Y9}),
$q=k+\alpha p$. Then
\begin{eqnarray}
\Gamma_{\sigma_1\sigma_1}(p)
&=&i\int_0^1d\alpha \int\frac{d^3q}{(2\pi)^3}\frac{-4\widetilde p^2(1-\alpha)-4\widetilde q\widetilde p+8\langle\sigma_1\rangle^2}{[\widetilde q^2+\widetilde p^2\alpha (1-\alpha)-\langle\sigma_1\rangle^2]^2}\nonumber\\
&=&i\int_0^1d\alpha \int\frac{d^3q}{(2\pi)^3}\frac{-4\widetilde p^2(1-\alpha)+8\langle\sigma_1\rangle^2}{[\widetilde q^2+\widetilde p^2\alpha (1-\alpha)-\langle\sigma_1\rangle^2]^2}.\label{Y10}
\end{eqnarray}
Note that in the second line of this equation we have ignored in the
numerator  of the fraction the linear term in  $\widetilde q$,
which evidently does not contribute to the $q$ integration in eq. (\ref{Y10}).

Suppose now that $\widetilde p^2<0$ in eq. (\ref{Y10}).
In this case one can perform in eq. (\ref{Y10}) a
Wick rotation of the $q_0$-integration contour and change
variables there, $q_0\to iq_0$, $q_1\to q_1/v_{\rm F}$ and $q_2\to
q_2/v_{\rm F}$. As a result, we obtain the integration over 3-dim
Euclidean $q$-momentum space. Using in the $q$-integral the polar
coordinate system, where $\int d^3q=4\pi\int x^2dx$ and
$x=\sqrt{q_0^2+q_1^2+q_2^2}$, we have
\begin{eqnarray}
\Gamma_{\sigma_1\sigma_1}(p)
&=&-\frac{1}{v_{\rm F}^2}\int_0^1d\alpha [-4\widetilde
p^2(1-\alpha)+8\langle\sigma_1\rangle^2]
\int_0^\infty\frac{dx}{2\pi^2}
\frac{x^2}{[x^2+\mu^2]^2},\label{Y11}
\end{eqnarray}
where $\mu^2=-\widetilde p^2\alpha
(1-\alpha)+\langle\sigma_1\rangle^2$. The integration over $x$ in eq.
(\ref{Y11}) is trivial, so
\begin{eqnarray}
\Gamma_{\sigma_1\sigma_1}(p) &=&-\frac{1}{8\pi v_{\rm
F}^2}\int_0^1d\alpha \frac{-4\widetilde
p^2(1-\alpha)+8\langle\sigma_1\rangle^2}{\sqrt{-\widetilde
p^2\alpha (1-\alpha)+\langle\sigma_1\rangle^2}} .\label{Y13}
\end{eqnarray}
To integrate in eq. (\ref{Y13}) one can use the substitution
$\alpha=\beta+1/2$. Note that in the obtained integral we can ignore
again in the numerator of the fraction the linear over $\beta$
term, so
\begin{eqnarray}
\Gamma_{\sigma_1\sigma_1}(p) &=&-\frac{1}{8\pi v_{\rm
F}^2}\int_{-1/2}^{1/2}d\beta \frac{-2\widetilde
p^2+8\langle\sigma_1\rangle^2}{\sqrt{-\widetilde
p^2}\sqrt{a^2-\beta^2}},\label{Y14}
\end{eqnarray}
where $a^2=\left [\langle\sigma_1\rangle^2-\widetilde
p^2/4\right ]/(-\widetilde
p^2)=1/4-\langle\sigma_1\rangle^2/\widetilde p^2$ (Note,
$a^2>1/4$). Hence,
\begin{eqnarray}
\Gamma_{\sigma_1\sigma_1}(p) &=&
\frac{\widetilde p^2-4\langle\sigma_1\rangle^2}{4\pi
v_{\rm F}^2\sqrt{-\widetilde p^2}}\arcsin\left
(\frac{\beta}{a}\right )\Bigg |^{\beta =1/2}_{\beta =-1/2}=
\frac{\widetilde p^2-4\langle\sigma_1\rangle^2}{2\pi
v_{\rm F}^2\sqrt{-\widetilde p^2}}\arcsin\left
(\frac{1}{2a}\right ).\label{Y15}
\end{eqnarray}
Finally, in order to obtain the expression (\ref{3.19}) for $\Gamma_{\sigma_1\sigma_1}(p)$, it is necessary to use in eq. (\ref{Y15}) the identity
\begin{eqnarray}
\arcsin x=\arctan\left (\frac{x}{\sqrt{1-x^2}}\right ).\label{Y16}
\end{eqnarray}

In a similar way one can obtain the expressions for the other 1PI Green functions of eq. (\ref{3.19}). For example, to find
$\Gamma_{\varphi_1\varphi_1}(p)$, we should use in eq. (\ref{3.17})
the substitution $\hat t_k =i\gamma^5$. Then, by analogy with
eq. (\ref{Y4}), we have
\begin{eqnarray}
\Gamma_{\varphi_1\varphi_1}(p)=-\left (\frac{1}{2v_{\rm F}H_{1}}-\frac{1}{2v_{\rm F}G_{1}}\right )-\frac{1}{2v_{\rm F}G_{1}}-i \int\frac{d^3k}{(2\pi)^3}{\rm
Tr}_{s}\left [\left
(\frac{1}{\lefteqn{/}\widetilde k+\lefteqn{/}\widetilde p-\langle\sigma_1\rangle }\right
)\gamma^5\left
(\frac{1}{\lefteqn{/}\widetilde k-\langle\sigma_1\rangle }\right
)\gamma^5\right ].\label{Y17}
\end{eqnarray}
Due to the relations (\ref{3.10}) for bare coupling constants $G_1$
and $H_1$, the expression in the first round brackets in
eq. (\ref{Y17}) is equal to $(h_1-g_1)/2v_{\rm F}$.
Taking into account eq. (\ref{Y5}), the sum of other terms in eq. (\ref{Y17}) brings us to the following expression
\begin{eqnarray}
\Gamma_{\varphi_1\varphi_1}(p)&=&-\frac{h_1-g_1}{2v_{\rm F}}+i \int\frac{d^3k}{(2\pi)^3}\left [\frac{-4\widetilde p^2-4\widetilde k\widetilde p}{(\widetilde k^2-\langle\sigma_1\rangle^2)((\widetilde k+\widetilde p)^2-\langle\sigma_1\rangle^2)}\right ].\label{Y18}
\end{eqnarray}
Applying in eq. (\ref{Y18}) the $\alpha$-representation formula (\ref{Y8}), we obtain after several variable changes both in the $k$-integration, $k=q-\alpha p$, and then in the $\alpha$-integration, $\alpha=\beta+1/2$, the following expression
\begin{eqnarray}
\Gamma_{\varphi_1\varphi_1}(p)&=&-\frac{h_1-g_1}{2v_{\rm F}}-2i\widetilde p^2\int_{\beta =-1/2}^{\beta=1/2}d\beta \int\frac{d^3q}{(2\pi)^3} \frac{1}{[\widetilde q^2+\widetilde p^2(1/4-\beta^2)-\langle\sigma_1\rangle^2]^2}.\label{Y19}
\end{eqnarray}
It can be evaluated by using the Wick-rotation
technique, which results in two table integrations both over $q$ and $\beta$ (similar calculations are presented after eq. (\ref{Y10})). As a result, we obtain the 2-point 1PI Green function of the $\varphi_1$ fields (\ref{3.19}).

In a similar way it is possible to get the expressions (\ref{3.19}) for
$\Gamma_{\varphi_2\varphi_2}(p)$ and $\Gamma_{\sigma_2\sigma_2}(p)$.

\section*{References}
\bibliography{EKKZ}

\end{document}